\documentclass[12pt,a4paper]{article}
\usepackage{amsfonts}
\usepackage{amsmath}
\usepackage{amssymb}
\usepackage{latexsym}
\numberwithin{equation}{section}

\newcommand{\lanln}[1]{$\langle$\texttt{arXiv:#1}$\rangle$}
\newcommand{\SLtwor}{{\mathrm{SL}}(2,\mathbb{R})}
\newcommand{\sltwor}{{\mathfrak{sl}}(2,\mathbb{R})}
\newcommand{\BbbR}{\mathbb{R}}
\newcommand{\BbbZ}{\mathbb{Z}}
\newcommand{\be}{\begin{eqnarray}}
\newcommand{\ee}{\end{eqnarray}}
\newcommand{\nn}{\nonumber}
\newcommand{\com}[2]{\mbox{$[#1,#2]$}}

\newtheorem{proposition}{Proposition}[section]
\newtheorem{theorem}[proposition]{Theorem}

\newcommand{\myproof}{\emph{Proof\/}. }
\newcommand{\myremark}{\emph{Remark\/}. }

\newcommand{\Aclass}{{\mathcal{A}_\mathrm{class}}}
\newcommand{\Vphys}{{\mathcal{V}_\mathrm{phys}}}

\newcommand{\Aphystar}{\mathcal{A}^{(\star)}_\mathrm{phys}}

\newcommand{\Haux}{{\mathcal{H}_\mathrm{aux}}}
\newcommand{\Hauxplusplus}{{\mathcal{H}_\mathrm{aux}^{++}}}

\newcommand{\Hraq}{{\mathcal{H}_\mathrm{RAQ}}}

\newcommand{\Aobs}{{\mathcal{A}_\mathrm{obs}}}

\newcommand{\qnought}{\textsf{q}_0}

\newcommand{\barGamma}{{\overline\Gamma}}
\newcommand{\barGammanought}{\barGamma_0}
\newcommand{\barGammareg}{\barGamma_{\mathrm{reg}}}
\newcommand{\barGammaex}{\barGamma_{\mathrm{ex}}}

\newcommand{\Mred}{\mathcal{M}}
\newcommand{\Mnought}{\mathcal{M}_0}
\newcommand{\Mreg}{\mathcal{M}_{\mathrm{reg}}}
\newcommand{\Mex}{\mathcal{M}_{\mathrm{ex}}}

\newcommand{\Nreg}{\mathcal{N}_{\mathrm{reg}}}

\newcommand{\wmom}{\varpi}
\newcommand{\wmomvec}{\boldsymbol{\wmom}}

\newcommand{\Gext}{G_\mathrm{ext}}

\title{Refined algebraic quantisation 
with the 
\\
triangular subgroup of $\SLtwor$}

\author{Jorma Louko\thanks{jorma.louko@nottingham.ac.uk}
\ and
Alberto Molgado\thanks{pmxam@nottingham.ac.uk}
\\
\noalign{\vspace{3ex}}
\small{\it School of Mathematical Sciences,
University of Nottingham,}\\
\small{\it Nottingham NG7 2RD, UK}
\\
\noalign{\vspace{1ex}}\\
\small{(Revised February 2005)} 
\\
\noalign{\vspace{1ex}}
\small{\lanln{gr-qc/0404100}}
\\
\noalign{\vspace{1ex}}
\small{Published in International Journal of Modern Physics D 
{\bf 14}, 1131--1157
(2005).}
}

\date{}

\begin{document}


\maketitle

\begin{abstract}
We investigate refined algebraic quantisation with group averaging in
a constrained Hamiltonian system whose gauge group is the connected
component of the lower triangular subgroup of $\SLtwor$. The unreduced
phase space is $T^*\BbbR^{p+q}$ with $p\ge1$ and $q\ge1$, and the
system has a distinguished classical $\mathfrak{o}(p,q)$ observable
algebra.  Group averaging with the geometric average of the right and
left invariant measures, invariant under the group inverse, yields a
Hilbert space that carries a maximally degenerate principal unitary
series representation of ${\mathrm{O}}(p,q)$. The representation is
nontrivial iff $(p,q) \neq (1,1)$, which is also the condition for the
classical reduced phase space to be a symplectic manifold up to a
singular subset of measure zero. We present a detailed comparison to
an algebraic quantisation that imposes the constraints in the sense
$\hat{H}_a
\Psi =0$ and postulates self-adjointness of the $\mathfrak{o}(p,q)$
observables. Under certain technical assumptions that parallel those
of the group averaging theory, this algebraic quantisation gives no
quantum theory when $(p,q) = (1,2)$ or $(2,1)$, or when $p\ge2$,
$q\ge2$ and $p+q
\equiv 1
\pmod2$. 
\end{abstract}

\newpage

\section{Introduction}
\label{sec:intro} 

In quantisation of constrained systems, an elegant idea for
constructing the physical Hilbert space is to average states in an
auxiliary Hilbert space $\Haux$ over a suitable action of the gauge
group
\cite{AH,KL,QORD,epistle,BC,lands-against,lands-wren,%
GM2,GoMa,LouRov,Giulini-rev,Marolf-MG,Shvedov,LouMol}.  For a
noncompact gauge group the averaging need not converge on all of
$\Haux$, but when the averaging is formulated within refined algebraic
quantisation \cite{epistle,GM2,Marolf-MG}, convergence on a suitable
linear subspace will suffice, and such convergence has been found to
occur in concrete examples \cite{GoMa,LouRov,LouMol}. Results on the
equivalence of refined algebraic quantisation with other methods 
\cite{Shvedov,GM1} further show that
group averaging provides considerable control over the quantisation.

In this paper we study group averaging in a system with a
nonunimodular gauge group. The interest of
this situation arises from the rather different senses in which group
averaging satisfies the Dirac constraints for unimodular and
nonunimodular groups~\cite{GM2}. To guarantee that the would-be
inner product provided by group averaging is real, the averaging
measure needs to be invariant under the group inverse. For a
unimodular group, the left and right invariant Haar measure has this
property. For a nonunimodular group, the left and right invariant Haar
measures do not coincide, and neither is invariant under the group
inverse, but their geometric average $d_0g$ is. Suppose $\hat{H}_a$
are the constraint operators that generate the unitary gauge group
action on the auxiliary Hilbert space, with the commutators
\begin{equation}
[ \hat{H}_a \, , \hat{H}_b ]   = i C^c{}_{ab} \hat{H}_c
\ \ , 
\end{equation}
where 
$C^c{}_{ab}$ are the structure constants of the Lie algebra. 
As was shown in \cite{GM2} and will be reviewed in section
\ref{sec:RAQoutline} below, 
group averaging with $d_0g$ gives physical states $\Psi$ that satisfy
\begin{equation}
\hat{H}_a \Psi = - \frac{i}{2} C^b{}_{ab} \Psi
\ \ , 
\label{eq:amm-Dirac}
\end{equation} 
which agrees with the na\"\i{}ve Dirac quantum constraints, $\hat{H}_a
\Psi =0$, if and only if the group is unimodular. The term on the
right-hand side of (\ref{eq:amm-Dirac}) leads to no known
inconsistencies: For systems amenable to geometric quantisation in
both reduced and unreduced phase space, (\ref{eq:amm-Dirac}) is in
fact the form of Dirac constraints equivalent to reduced phase space
quantisation~\cite{Tuyn,DEGT,DEGST}.
Related observations for first class constrained systems with one
constraint quadratic in the momenta and several constraints linear in
the momenta were made in~\cite{HaKu}. 

We shall consider a system obtained by replacing a unimodular gauge
group $G_u$ by its nonunimodular subgroup~$G$. The effect of this
replacement on the
constraints 
$\hat{H}_a \Psi =0$ of $G_u$ 
is not just that some of the
constraint equations are dropped: 
In the constraints that remain, 
(\ref{eq:amm-Dirac}) shows that new 
terms appear on the right-hand side. 
An observable algebra that
commutes strongly with the old constraints is still represented on the
solution space to the new constraints, but the new representation need
not be isomorphic to the old representation. In group averaging, where
no observables need be explicitly constructed, these changes are
encoded in the integration measures on $G_u$ and~$G$. The integrals
over $G_u$ and $G$ may also have differing convergence properties, and
it may hence be necessary to choose the test spaces differently even
when $\Haux$ is unchanged and the representation of $G$ is obtained
from that of $G_u$ by restriction. Our system will indeed exemplify
all these phenomena. 

Our unimodular gauge system \cite{LouMol} has gauge group $G_u
\simeq\SLtwor$, unreduced phase space $T^*\BbbR^{p+q} \simeq
\BbbR^{2(p+q)}$ with $p\ge1$ and $q\ge1$ and a distinguished classical
$\mathfrak{o}(p,q)$ observable algebra. The reduced phase space is a
symplectic manifold up to a singular subset of measure zero if and
only if $p\ge2$ and $q\ge2$: This manifold has dimension $2(p+q-3)$
and is separated by the $\mathfrak{o}(p,q)$ observables. The system
was quantised with group averaging in
\cite{LouRov,LouMol}, recovering a quantum theory with a nontrivial
representation of the quantum $\mathfrak{o}(p,q)$ observables when 
$p\ge2$, $q\ge2$ and $p+q \equiv 0 \pmod2$. The quadratic
$\mathfrak{o}(p,q)$ Casimir was
found to take the value $-\frac14 (p+q)(p+q-4)$. 
Quantisations of this system for special values of $p$ and $q$ by a
variety of other methods can be found in
\cite{LouRov,LouMol,MRT,Monte,GamPor,TruSL,bojo-etal,bojo-strobl,%
kastrup,BarsKounnas,Bars98,Bars00,Bars01,Dittrich-sl2r}.

Our nonunimodular gauge group $G \subset G_u$ is the connected
component of the lower triangular subgroup of $\SLtwor$. $G$~is
two-dimensional and nonabelian, and hence isomorphic to every
two-dimensional connected nonabelian group. The reduced phase space is
a symplectic manifold up to a singular subset of measure zero if and
only if $(p,q)
\ne (1,1)$: This manifold is symplectomorphic to $T^*(S^{p-1} \times
S^{q-1})$, and it is separated by the $\mathfrak{o}(p,q)$ observables
up to a set of measure zero. Quantisation with group averaging gives a
quantum theory with a nontrivial representation of the quantum
$\mathfrak{o}(p,q)$ observables for all $(p,q) \ne (1,1)$. This
representation is the end-point of the maximally degenerate principal
unitary series~\cite{Vil}, and the quadratic Casimir takes the value
$-\frac14 {(p+q-2)}^2$.

For comparison, we quantise the system also with the constraints
$\hat{H}_a
\Psi =0$, adopting the algebraic quantisation
framework of \cite{Ash1,Ash2} and requiring the $\mathfrak{o}(p,q)$
observables to become self-adjoint operators. For $(p,q) = (1,3)$,
$(3,1)$ and $(2,2)$, the group averaging theory and the algebraic
quantisation theory are qualitatively fairly similar, with minor
differences in the spectra of certain operators. Our algebraic
quantisation results for general $(p,q)$ remain incomplete, but we
show that under certain technical assumptions that parallel those of
the group averaging theory, the algebraic quantisation gives no
quantum theory for $(p,q) = (1,2)$ and $(2,1)$, or for $p\ge2$,
$q\ge2$ and $p+q
\equiv 1
\pmod2$. 

The rest of the paper is as follows. Section \ref{sec:classical}
introduces and analyses the classical system. Sections \ref{sec:AQ}
and \ref{sec:AQnaive} discuss algebraic quantisation, with
respectively the constraints (\ref{eq:amm-Dirac}) and $\hat{H}_a
\Psi =0$. Refined algebraic
quantisation with group averaging is briefly reviewed in section
\ref{sec:RAQoutline}
and carried out in section 
\ref{sec:RAQus} for $(p,q) \ne (1,1)$ and in section 
\ref{sec:RAQ-low} for $p=q=1$. 
Section \ref{sec:discussion} presents a summary and concluding
remarks. 

Appendix \ref{app:affine} collects some basic properties of the
group~$G$. The separation properties of the $\mathfrak{o}(p,q)$
observables on the reduced phase space are verified in
appendix~\ref{app:separation}, and certain technical results for refined
algebraic quantisation with $(p,q) = (1,1)$ are proved in
appendix~\ref{app:dense}.

\section{Classical system}
\label{sec:classical}

In this section we introduce a classical constrained system with the
unreduced phase space $T^*\mathbb{R}^{p+q}$, where $p\ge1$ and
$q\ge1$. The system is obtained from the $\SLtwor$ system of
\cite{LouMol} by dropping the constraint $H_1$ therein and relabelling one
of the canonical pairs by $(\boldsymbol{v}, \boldsymbol{\pi}) =
(-\wmomvec, \boldsymbol{w})$. The gauge transformations and the
$\mathfrak{o}(p,q)$ observables are therefore obtained directly
from~\cite{LouMol}.  We shall however see that the structure of the
reduced phase space differs markedly from that in~\cite{LouMol}.

\subsection{The system}
\label{subsec:model}

The system is defined by the action
\be
S=\int 
dt \, 
\bigl( \boldsymbol{p} \cdot \dot{\boldsymbol{u}}
+ \wmomvec  \cdot \dot{\boldsymbol{w}} 
-MH -\lambda D
\bigr)
\ \ ,
\label{eq:act1}
\ee
where $\boldsymbol{u}$ and 
$\boldsymbol{p}$ are real vectors of dimension~$p\ge1$,
$\boldsymbol{w}$ and $\wmomvec$ are real vectors of
dimension $q\ge1$ and the overdot denotes differentiation with respect
to~$t$. The symplectic structure is
\begin{equation}
\Omega = \sum_{i=1}^p dp_i \wedge du_i
+
\sum_{j=1}^q d\wmom_j \wedge dw_j
\ \ ,
\label{eq:Omega-on-Gamma}
\end{equation}
and the phase space is
$\Gamma:=T^*\mathbb{R}^{p+q} \simeq \mathbb{R}^{2(p+q)}$.
$M$ and $\lambda$ are Lagrange multipliers
associated with the constraints
\be
H &:=& \tfrac12 (\boldsymbol{w}^2-\boldsymbol{u}^2)
\ \ ,
\nn
\\
D &:=& \boldsymbol{u} \cdot \boldsymbol{p} + 
\boldsymbol{w} \cdot \wmomvec
\ \ .
\label{eq:const}
\ee
The Poisson algebra of
the constraints is 
\be
\{H\, , D\}  &=&2H
\ \ ,
\label{eq:algebra}
\ee
and the system is a first class constrained
system~\cite{Dir3, Hen}. 
The finite gauge transformations on $\Gamma$
generated by the constraints are
\be
\left(
\begin{array}{c}
\boldsymbol{u}\\
\boldsymbol{p}\\
\end{array}
\right)\mapsto  g\left(
\begin{array}{c}
\boldsymbol{u}\\
\boldsymbol{p}\\
\end{array}
\right), \hspace{5ex}
\left(
\begin{array}{c}
\boldsymbol{w}\\
\wmomvec\\
\end{array}
\right)\mapsto  
\left(
\begin{array}{cc}
1 & 0
\\
0 & -1
\\
\end{array}
\right) 
g 
\left(
\begin{array}{cc}
1 & 0
\\
0 & -1
\\
\end{array}
\right) 
\left(
\begin{array}{c}
\boldsymbol{w}\\
\wmomvec\\
\end{array}
\right),
\label{eq:gauge-transf}
\ee
where $g \in \SLtwor$ is a lower diagonal matrix with positive
diagonal elements. Relevant properties of the gauge group $G$
and its Lie algebra are collected in appendix~\ref{app:affine}. 


\subsection{Classical observables}
\label{subsec:class-obs}

Recall that an observable is a function on $\Gamma$ whose
Poisson brackets with the first class
constraints vanish when the first class constraints hold~\cite{Hen}.
As discussed in~\cite{LouMol}, the system has the observables 
\begin{equation}
\begin{array}{lcll}
A_{ij}
& := &
u_i p_j-u_j p_i
\ ,
\hspace{7ex}
&
1\le i \le p,
\ \
1\le j\le p
\ ;
\\
B_{ij}
& := &
w_i \wmom_j-w_j \wmom_i
\ ,
&
1\le i \le q,
\ \
1 \le j\le q
\ ;
\\
C_{ij}
& := &
- u_i \wmom_j
- p_i w_j
\ ,
&
1\leq i\leq p,
\ \
1\leq j\leq q
\ ,
\end{array}
\label{eq:obser}
\end{equation}
whose Poisson brackets close as the $\mathfrak{o}(p,q)$ Lie algebra,
$\{A_{ij}\}$ and $\{B_{ij}\}$ spanning respectively the
$\mathfrak{o}(p)$ and $\mathfrak{o}(q)$ Lie subalgebras. We denote the
algebra generated by the observables (\ref{eq:obser})
by~$\Aclass$. The finite transformations that $\Aclass$ generates on
$\Gamma$ are
\be
\left(
\begin{array}{c}
\boldsymbol{u}\\
\boldsymbol{w}\\
\end{array}
\right)\mapsto  R
\left(
\begin{array}{c}
\boldsymbol{u}\\
\boldsymbol{w}\\
\end{array}
\right), \hspace{5ex}
\left(
\begin{array}{c}
\boldsymbol{p}\\
\wmomvec\\
\end{array}
\right)\mapsto  {(R^T)}^{-1}
\left(
\begin{array}{c}
\boldsymbol{p}\\
\wmomvec\\
\end{array}
\right),
\label{eq:alg-action}
\ee
where $R$ is a matrix in the connected component 
$\mathrm{O}_{\mathrm{c}}(p,q)$ of $\mathrm{O}(p,q)$.

The quadratic Casimir element in $\Aclass$ is \cite{AMMP}
\begin{equation}
\mathcal{C}
:= 
\sum_{i < j} {(A_{ij})^2}
\
+
\sum_{i < j} {(B_{ij})^2}
\
-
\sum_{i,j} {(C_{ij})^2}
\ \ . 
\label{eq:class-casimir}
\end{equation}
When the constraints hold, $\mathcal{C}$ vanishes~\cite{LouMol}.

\subsection{Reduced phase space}
\label{subsec:red-phasespace}

Let $\barGamma$ be the subset of $\Gamma$ where the constraints
hold. The reduced phase space, denoted by~$\Mred$, is the
quotient of $\barGamma$ under the gauge action~(\ref{eq:gauge-transf}).
We now discuss the structure of~$\Mred$. 

Note first that as the Hamiltonian is a linear combination of the
constraints, there is no dynamics on~$\Mred$, and $\Mred$ can be
identified with the space of classical solutions. As the functions in
$\Aclass$ are gauge invariant, they project to functions on~$\Mred$.

$\barGamma$ is clearly connected.
Hence also $\Mred$ is connected.

Let $\barGammanought = \{ \qnought \}$, where $\qnought$ is the origin
of~$\Gamma$, $\boldsymbol{u} = \boldsymbol{p} = \boldsymbol{0} =
\boldsymbol{w} =
\wmomvec$. 
Let $\barGammaex$ contain all other points on $\barGamma$ at which
$\boldsymbol{u} = \boldsymbol{w} = \boldsymbol{0}$, and let
$\barGammareg$ contain the points on $\barGamma$ at which
$\boldsymbol{u} \ne
\boldsymbol{0} \ne \boldsymbol{w}$. By the constraint $H=0$, 
$\barGamma = \barGammanought \cup \barGammaex \cup \barGammareg$. We
refer to $\barGammaex$ and $\barGammareg$ as respectively the
``exceptional'' and ``regular'' parts of~$\barGamma$.

As $\barGammanought$, $\barGammaex$ and $\barGammareg$ are preserved
by the gauge transformations, they project onto disjoint subsets
of~$\Mred$. We denote these sets respectively by $\Mnought$, $\Mex$
and $\Mreg$ and analyse each in turn.

\subsubsection{$\Mnought$}

$\Mnought$ contains only one point, the projection of~$\qnought$.
All observables in $\Aclass$ vanish on~$\Mnought$.

\subsubsection{$\Mex$}

Each point in $\barGammaex$ is gauge-equivalent to a unique point that
satisfies
\begin{equation}
\boldsymbol{u} = \boldsymbol{0} = \boldsymbol{w}
\ \ ,
\ \
\boldsymbol{p}^2 + \wmomvec^2 =1
\ \ .
\label{eq:Mex-char}
\end{equation}
$\Mex$ has thus topology~$S^{p+q-1}$. By~(\ref{eq:Omega-on-Gamma}),
the projection of $\Omega$ vanishes on~$\Mex$. All observables in
$\Aclass$ vanish on~$\Mex$.

\subsubsection{$\Mreg$}

Each point in $\barGammareg$ is gauge-equivalent to a unique point
that satisfies
\begin{equation}
\boldsymbol{u}^2 = \boldsymbol{w}^2 = 1
\ \ ,
\ \
\boldsymbol{u} \cdot \boldsymbol{p} 
= \boldsymbol{w} \cdot \wmomvec
= 0 
\ \ .
\label{eq:Mreg-char}
\end{equation}
$\Mreg$ can therefore be represented as the set~(\ref{eq:Mreg-char}),
which is the cotangent bundle over $S^{p-1}
\times S^{q-1}$, with 
$(\boldsymbol{u},\boldsymbol{w})$ forming the base space and
$(\boldsymbol{p}, \wmomvec)$ the fibres.
By~(\ref{eq:Omega-on-Gamma}), the symplectic structure on $\Mreg$ induced
from $\Gamma$ is precisely the symplectic structure of this cotangent bundle
description.
$\Mreg$~is connected when 
$p>1$ and $q>1$, and it has two connected components when exactly one of
$p$ and $q$ equals~$1$. When $p=q=1$, $\Mreg$ 
contains just four points. 

$\Aclass$ does not separate all of~$\Mreg$. However, we show in
appendix
\ref{app:separation} that when $p>1$ and $q>1$, $\Aclass$ separates the
subset of $\Mreg$ in which $0 \ne \boldsymbol{p}^2 \ne \wmomvec^2
\ne 0$, in the gauge~(\ref{eq:Mreg-char}), up to the twofold degeneracy 
\begin{equation}
(\boldsymbol{u}, \boldsymbol{w}, \boldsymbol{p}, \wmomvec)
\mapsto  
(-\boldsymbol{u}, -\boldsymbol{w}, -\boldsymbol{p}, -\wmomvec)
\ \ . 
\label{eq:separ-degen}
\end{equation} 
We also show that when $p=1$ and $q>1$ (respectively $p>1$ and $q=1$),
$\Aclass$ separates the subset of $\Mreg$ in which $\wmomvec^2 \ne 0$
($\boldsymbol{p}^2 \ne 0$), again up to the
degeneracy~(\ref{eq:separ-degen}). When $p=q=1$, all observables in
$\Aclass$ vanish on~$\Mreg$.

\section{Algebraic quantisation with the 
constraints~(\ref{eq:amm-Dirac})}
\label{sec:AQ}

In this section we quantise the system by the algebraic techniques
of~\cite{Ash1,Ash2}, imposing the constraints in the
form~(\ref{eq:amm-Dirac}), promoting the classical $\mathfrak{o}(p,q)$
observables into an $\mathfrak{o}(p,q)$ algebra of operators, and
seeking an inner product in which these quantum observables are
self-adjoint. Subsection
\ref{subsec:AQ-genpq} shows that this procedure leads to quantum
theories for arbitrary $(p,q)$: It will be seen in sections
\ref{sec:RAQus} and \ref{sec:RAQ-low} 
that one of these quantum theories is isomorphic to that emerging from
refined algebraic quantisation with group averaging. The remaining
subsections analyse the quantum theories in detail for $p+q \le4$.
The use of the classical $\mathfrak{o}(p,q)$
observables is motivated by their 
separation properties on the reduced phase space 
(see subsection 
\ref{subsec:red-phasespace} above).

\subsection{General $(p,q)$}
\label{subsec:AQ-genpq}

Treating $\boldsymbol{u}$ and $\boldsymbol{w}$ as the `configuration'
variables, we represent the elementary operators as 
\be
\hat{\boldsymbol{u}}\Psi
=
\boldsymbol{u}\Psi
\ ,
\hspace{5ex}
\hat{\boldsymbol{p}} \, \Psi
=
-i\boldsymbol{\nabla}_u \Psi
\ ,
\nn
\\
\hat{\boldsymbol{w}} \, \Psi 
=
\boldsymbol{w}\Psi
\ ,
\hspace{5ex}
\hat{\boldsymbol{\wmomvec}} \Psi
=
-i\boldsymbol{\nabla}_w \Psi 
\ , 
\label{eq:operator}
\ee
where the class of `functions' 
$\Psi(\boldsymbol{u},\boldsymbol{w})$ will be specified shortly. 
For the quantum constraint operators, we take 
\begin{subequations}
\label{eq:con-op}
\be
\hat{H} & := & \tfrac12 \! 
\left( \boldsymbol{w}^2  - \boldsymbol{u}^2 \right)
\ \ ,
\\
\hat{D} & := &
- i \left(
\boldsymbol{u}\cdot\boldsymbol{\nabla}_u
+ \boldsymbol{w}\cdot\boldsymbol{\nabla}_w
+
\frac{p+q}{2}
\right)
\ \ , 
\label{eq:con-op-D}
\ee
\end{subequations}
so that the commutator algebra reads 
\begin{equation}
\com{\hat{H}}{\hat{D}} = 2i\hat{H}
\ \ .
\label{eq:q-commutator}
\end{equation}
The constraints (\ref{eq:con-op}) are symmetric in the inner
product in which the integration measure is $d^p \boldsymbol{u} \, d^q
\boldsymbol{w}$, 
as motivated by comparison with 
refined algebraic quantisation in sections \ref{sec:RAQus}
and~\ref{sec:RAQ-low}. Note however that algebraic quantisation does
not introduce an inner product at this stage, and the physical inner
product will not be integration in the measure $d^p
\boldsymbol{u} 
\, d^q
\boldsymbol{w}$. 

The quantum constraints (\ref{eq:amm-Dirac}) read 
\begin{subequations}
\label{eq:qconstraints}
\be
&&\hat{H} \Psi =0
\ \ ,
\label{eq:qconstraints-H}
\\
&&
(\hat{D} - i) \Psi =0
\ \ . 
\label{eq:qconstraints-D}
\ee
\end{subequations}
As the only continuous solution to (\ref{eq:qconstraints-H}) is
$\Psi=0$, we seek solutions in the space of (say, Schwarz)
distributions with the integration measure $d^p \boldsymbol{u} \, d^q
\boldsymbol{w}$. Equation (\ref{eq:qconstraints-D}) is equivalent to
the homogeneity condition
$\Psi(r \boldsymbol{u}, r \boldsymbol{w}) 
= r^{-(p+q+2)/2} \Psi
(\boldsymbol{u}, \boldsymbol{w})$ for $r>0$. The set
(\ref{eq:qconstraints}) is thus satisfied by 
\begin{subequations}
\label{eq:Vphys}
\begin{equation}
\Psi (\boldsymbol{u}, \boldsymbol{w}) = \delta 
(u^2 - w^2) 
f (\boldsymbol{u}, \boldsymbol{w})
\ \ , 
\end{equation}
where $u := \sqrt{\boldsymbol{u}^2}$ and $w :=
\sqrt{\boldsymbol{w}^2}$, $\delta$ is the Dirac delta-distribution,
and $f(\boldsymbol{u}, \boldsymbol{w})$ is smooth for
$(\boldsymbol{u}, \boldsymbol{w}) 
\ne 
(\boldsymbol{0}, \boldsymbol{0})$ 
and homogeneous of degree $-(p+q-2)/2$, 
\begin{equation}
f (r \boldsymbol{u}, r \boldsymbol{w}) 
= r^{-(p+q-2)/2} f
(\boldsymbol{u}, \boldsymbol{w})
\ \ \mathrm{for} \ r>0
\ \ . 
\label{eq:f-homog}
\end{equation} 
\end{subequations}
We denote the vector space of the 
solutions (\ref{eq:Vphys}) by~$\Vphys$. 

We define the quantum counterparts of the classical
observables (\ref{eq:obser}) as 
\begin{equation}
\begin{array}{lcll}
\hat{A}_{ij}
& := &
-i \left( u_i \partial_{u_j} - u_j \partial_{u_i} \right)
\ ,
\hspace{7ex}
&
1\le i \le p,
\ \
1\le j \le p
\ ;
\\
\hat{B}_{ij}
& := &
-i \left( w_i \partial_{w_j} - w_j  \partial_{w_i} \right)
\ ,
&
1\le i \le q,
\ \
1 \le j \le q
\ ;
\\
\hat{C}_{ij}
& := &
i \left( u_i \partial_{w_j} + w_j  \partial_{u_i} \right)
\ ,
&
1\leq i\leq p,
\ \
1\leq j\leq q
\ , 
\end{array}
\label{eq:q-obser}
\end{equation}
These operators commute with the quantum constraints (\ref{eq:con-op})
and are thus quantum observables, and their commutator algebra closes
as ($i$~times) the $\mathfrak{o}(p,q)$ Lie algebra. As
in~\cite{LouMol}, we define the full star-algebra $\Aphystar$ as the
algebra generated by~(\ref{eq:q-obser}), the antilinear star-relation
being defined so that it leaves the observables (\ref{eq:q-obser})
invariant. The quantum quadratic Casimir observable is \cite{AMMP}
\begin{equation}
\hat{\mathcal{C}}
:=
\sum_{i < j} {(\hat{A}_{ij})^2}
\
+
\sum_{i < j}{(\hat{B}_{ij})^2}
\
-
\sum_{i,j} {(\hat{C}_{ij})^2}
\ \ . 
\label{eq:pq-quantum-casimir}
\end{equation}
On states satisfying~(\ref{eq:qconstraints}), 
$\hat{\mathcal{C}}$ takes the
value $-\frac14 {(p+q -2)}^2$ \cite{LouMol}. 

Recall from Section 9.2.9 
of \cite{Vil} 
that $\Vphys$ carries a representation 
$T$ of $\mathrm{O}_{\mathrm{c}}(p,q)$ given by
\begin{equation}
[T(R)\Psi](\boldsymbol{u},
\boldsymbol{w})
= 
\Psi\bigl(R^{-1}(\boldsymbol{u},
\boldsymbol{w})\bigr)
\ \ , 
\end{equation}
where $R\in
\mathrm{O}_{\mathrm{c}}(p,q)$ acts on the configuration space as
in~(\ref{eq:alg-action}). Writing $\Psi = \delta f$ as
in~(\ref{eq:Vphys}), and noting that $u^2 - w^2$ is invariant under
$\mathrm{O}_{\mathrm{c}}(p,q)$, this representation reads 
\begin{equation}
[T(R) \delta f](\boldsymbol{u},
\boldsymbol{w})
= \delta (u^2 - w^2)
f \bigl(R^{-1}(\boldsymbol{u},
\boldsymbol{w})\bigr)
\ \ . 
\end{equation}
$T$~is thus isomorphic to the representation of
$\mathrm{O}_{\mathrm{c}}(p,q)$ on homogeneous functions of degree
$-(p+q-2)/2$ on the light cone of $\BbbR^{p,q}$. It was observed in 
\cite{Vil} that
$T$ is unitary in the inner product
\begin{equation} 
(\Psi_1, \Psi_2) = \int_{S^{p-1} \times S^{q-1}}
d\boldsymbol{u}\,d\boldsymbol{w} \,
\overline{f_1(\boldsymbol{u},\boldsymbol{w})} \, 
f_2(\boldsymbol{u},\boldsymbol{w}) 
\ \ , 
\label{eq:aq-ip} 
\end{equation}
where the overline denotes complex conjugation and the integration is
over the product of the unit spheres in $\boldsymbol{u}$
and~$\boldsymbol{w}$,
$\boldsymbol{u}^2=\boldsymbol{w}^2=1$. Completion of $\Vphys$ in the
inner product (\ref{eq:aq-ip}) therefore gives a physical Hilbert
space on which $T$ is unitary. 
The infinitesimal generators of $T$ 
are given by~(\ref{eq:q-obser}), and they are 
densely-defined self-adjoint operators on the physical Hilbert space. 

Now, in algebraic quantisation the task is to seek vector spaces on which 
the representation of $\Aphystar$ is (algebraically)
irreducible and on which there exists an inner product in which the
generators (\ref{eq:q-obser}) are self-adjoint. Each such subspace yields
a Hilbert space and an independent quantum theory by Cauchy
completion. The properties of $T$ show that each component 
in the decomposition of $T$ into unitary irreducible 
representations gives such a quantum theory, the vector 
space being a suitable domain of the infinitesimal 
generators of~$T$. In the following subsections we 
shall examine this issue in detail 
for $p+q \le4$, showing that 
an appropriate subspace of $\Vphys$ will provide 
the requisite vector space. 

We record here some properties of $T$~\cite{Vil}. For any $p$ and~$q$,
$T$ reduces into subrepresentations $T^\epsilon$ on functions of
parity $\epsilon \in \{0,1\}$, $f(-\boldsymbol{u},-\boldsymbol{w}) =
(-1)^\epsilon f(\boldsymbol{u},\boldsymbol{w})$. In the terminology
of~\cite{Vil}, $T^\epsilon$ is the end-point of the maximally
degenerate principal unitary series of representations of
$\mathrm{O}_{\mathrm{c}}(p,q)$, denoted by $T^{(p+q-2)/2,
\epsilon}_{pq}$. $T^{(p+q-2)/2, \epsilon}_{pq}$ is nontrivial iff $(p,q)
\ne (1,1)$. $T^{(p+q-2)/2, \epsilon}_{pq}$ is known to be 
irreducible when $p$ and $q$ are of opposite parity, and also when $p$
and $q$ are both even and $(p+q)/2 +\epsilon \equiv 0 \pmod2$.

\subsection{$(p,q) = (1,1)$}
\label{subsec:11-low} 

When $(p,q) = (1,1)$, $\Vphys$ has dimension four, corresponding to
the four branches of the 1+1 light cone. The
representation of $\Aphystar$ on $\Vphys$ is trivial and does not
restrict the choice of an inner product on~$\Vphys$.

\subsection{$(p,q) = (2,1)$}
\label{subsec:21-low} 

When $(p,q) = (2,1)$, we consider the subspaces ${\mathcal V}^\kappa :=
\mathrm{span} \{ \psi^\kappa_m \} \subset \Vphys$, 
where $\kappa \in \{1,-1\}$, 
\be
\psi^\kappa_m 
:= 
{(-i\kappa)}^m 
\delta(u^2 - w^2) 
\theta(\kappa w_1) 
u^{-1/2} 
e^{i m \alpha}
\ \ ,
\label{eq:psi-kappa-m}
\ee
$m \in \BbbZ$, $\theta$ is the Heaviside function, and we have written
$u_1 + iu_2 = u e^{i\alpha}$. Writing $\hat{C}_\pm := \hat{C}_{11} \pm
i \hat{C}_{21}$, the set $\{ \hat{A}_{12} , \hat{C}_+ , \hat{C}_- \}$
forms a standard raising and lowering operator basis for
$\mathfrak{o}(2,1)$
\cite{bargmann,lang}, and we find 
\begin{subequations}
\label{eq:21-rep}
\be
\hat{A}_{12} \psi^\kappa_m 
&=& 
m 
\psi^\kappa_m 
\ \ , 
\\
\hat{C}_\pm
\psi^\kappa_m 
&=& 
\bigl( m \pm \tfrac12 \bigr) 
\psi^\kappa_{m\pm1}
\ \ . 
\label{eq:21-rep-pm}
\ee
\end{subequations}
Each ${\mathcal V}^\kappa$ hence carries an irreducible representation
of $\Aphystar$. This means in particular that the branches $w_1>0$ and
$w_1<0$ of the light cone decouple. Requiring $\hat{A}_{12}$,
$\hat{C}_{11}$ and $\hat{C}_{21}$ to be self-adjoint determines the
inner product on each ${\mathcal V}^\kappa$ to be such that $\{
\psi^\kappa_m \}$ is an orthonormal set up to an overall
scale~\cite{bargmann}. Note that this agrees with the inner
product~(\ref{eq:aq-ip}). The Hilbert space is obtained by Cauchy
completion. The representation of $\mathrm{O}_{\mathrm{c}}(2,1)$ on
each of the two Hilbert spaces is isomorphic to the principal series
irreducible unitary representation denoted in
\cite{bargmann} by~$C^0_{1/4}$. The representations 
$T^{\frac12,\epsilon}_{2,1}$ arise by
starting from the parity $\epsilon$
subspaces $\mathrm{span} \{ \psi^{1}_m + {(-1)}^\epsilon \psi^{-1}_m
\} \subset {\mathcal V}^1 \oplus {\mathcal V}^{-1}$ and are each
isomorphic to~$C^0_{1/4}$. 

The spectra of the quantum observables in $\Aphystar$ are in
qualitative agreement with the ranges of the corresponding classical
observables. In particular, both $A_{12}$ and the spectrum of
$\hat{A}_{12}$ are unbounded above and below.

\subsection{$(p,q) = (2,2)$}
\label{subsec:22-low} 

When $(p,q) = (2,2)$, we write \cite{LouRov} 
\begin{eqnarray}
{\hat\tau}_0^\eta 
&:=& 
\tfrac12 \bigl( \hat{A}_{12} + \eta  \hat{B}_{12} \bigr)
\ \ ,
\nonumber
\\
{\hat\tau}_1^\eta 
&:=& 
\tfrac12 \bigl( \hat{C}_{11} - \eta  \hat{C}_{22}  \bigr)
\ \ ,
\nonumber
\\
{\hat\tau}_2^\eta 
&:=& 
\tfrac12 \bigl( \hat{C}_{21} + \eta  \hat{C}_{12}  \bigr)
\ \ ,
\label{tauhat-def}
\end{eqnarray}
where $\eta \in \{1,-1\}$
and 
\begin{equation}
{\hat\tau}_\pm^\eta
:= 
{\hat\tau}_1^\eta \pm i {\hat\tau}_2^\eta
\ \ .
\end{equation}
The commutators of the $\hat{\tau}$s are 
\begin{subequations}
\label{taus-qcomms}
\begin{eqnarray}
{\bigl[ {\hat\tau}_0^\eta  , \, {\hat\tau}_\pm^{\eta'} \bigr] }
& = & 
\pm \delta^{\eta \eta'} \, {\hat\tau}_\pm^\eta
\ \ ,
\\
{\bigl[  {\hat\tau}_+^\eta  , \, {\hat\tau}_-^{\eta'} \bigr] }
& = & 
- 2 \delta^{\eta \eta'} \, {\hat\tau}_0^\eta
\ \ , 
\end{eqnarray}
\end{subequations}
which shows that the $\hat{\tau}$s provide 
standard $\mathfrak{sl}(2,\BbbR)$ raising and
lowering operator bases 
in the decomposition 
$\mathfrak{o}(2,2) \simeq \mathfrak{sl}(2,\BbbR) \oplus
\mathfrak{sl}(2,\BbbR)$. 
In the polar coordinates $u_1
+ iu_2 = u e^{i\alpha}$, 
$w_1
+ iw_2 = w e^{i\beta}$, we find 
\begin{subequations}
\label{polartaus}
\begin{eqnarray}
{\hat\tau}_0^\eta
&=&
- \tfrac12 i 
\left( \partial_\alpha + \eta \partial_\beta \right)
\ \ ,
\\
{\hat\tau}_\pm^\eta
&=&
\tfrac{1}{2}i e^{\pm i(\alpha +
\eta\beta)} 
\bigl\{ 
u \bigl[\partial_w \pm \eta(i/w)\partial_\beta \bigr]
+ w \bigl[ \partial_u \pm (i/u)\partial_\alpha \bigr] 
\bigr\}
\ \ . 
\end{eqnarray}
\end{subequations}

Let $\mathcal{V} \subset \Vphys$ be the subspace spanned by the
vectors 
$\delta(u^2-w^2)u^{-1} e^{i(m\alpha+n\beta)}$, where $m$ and $n$
are integers. We label these vectors by $\mu
:=
\frac{1}{2}(m+n)$ and $\nu :=
\frac{1}{2}(m-n)$, defining 
\be
\psi_{\mu \nu}
:= {(-i)}^{\mu + \nu} 
\delta(u^2-w^2)u^{-1}
e^{i[(\mu+\nu)\alpha +(\mu-\nu)\beta]} 
\ \  ,   
\label{eq:psi22}
\ee
where $\mu$ and $\nu$ are either both integers or both half-integers.
A~direct computation gives
\be
{\hat\tau}_0^+ \psi_{\mu\nu}  
& = & 
\mu \psi_{\mu\nu} 
\ \ , 
\nonumber
\\
{\hat\tau}_\pm^+ \psi_{\mu\nu}  
& = & 
\bigl( \mu \pm \tfrac12 \bigr)
\psi_{\mu \pm 1, \nu} 
\ \ , 
\nonumber
\\
{\hat\tau}_0^{-} \psi_{\mu\nu}  
& = & 
\nu \psi_{\mu\nu} 
\ \ , 
\nonumber
\\
{\hat\tau}_\pm^{-} \psi_{\mu\nu}  
& = & 
\bigl( \nu \pm \tfrac12 \bigr)
\psi_{\mu, \nu  \pm 1} 
\ \ , 
\label{eq:repr22}
\ee
which shows that $\mathcal{V}$ carries a representation of~$\Aphystar$.

We decompose $\mathcal{V}$ as $\mathcal{V} = 
\mathcal{V}^{\mathrm{e}}
\oplus 
\bigoplus_{\epsilon_1 \epsilon_2}
\mathcal{V}^{\mathrm{o}}_{\epsilon_1 \epsilon_2}$, where 
\be
\mathcal{V}}^{\mathrm{e} 
&:=&
\mathrm{span}
\bigl\{ \psi_{\mu\nu} 
\mid 
\mu, \nu \in \BbbZ 
\bigr\}
\ \ ,
\nonumber
\\
\mathcal{V}^{\mathrm{o}}_{\epsilon_1 \epsilon_2}
&:=&
\mathrm{span}
\bigl\{ \psi_{\mu\nu} 
\mid 
\mu, \nu \in \BbbZ + \tfrac12 ; \ 
\mathrm{sgn}(\mu) = \epsilon_1 ; \ 
\mathrm{sgn}(\nu) = \epsilon_2
\bigr\}
\ \ ,
\label{eq:22Vdecomp} 
\ee
and $\epsilon_i \in \{1,-1\}$. Equations (\ref{eq:repr22}) show that
each of the five spaces in (\ref{eq:22Vdecomp}) carries a
representation of~$\Aphystar$. These representations are irreducible:
Given a nonzero vector in one of the spaces, acting on this vector
repeatedly with ${\hat\tau}_0^\eta$ and taking suitable linear combinations
generates some~$\psi_{\mu_0 \nu_0}$, and repeatedly acting on this
$\psi_{\mu_0
\nu_0}$ by
${\hat\tau}_\pm^\eta$ and taking linear combinations generates all of the
space.  Note that the difference between integer and half-integer
indices in (\ref{eq:22Vdecomp}) arises because in the latter case the
numerical coefficients in the raising and lowering operator action in
(\ref{eq:repr22}) may vanish.

On each of the five spaces in~(\ref{eq:22Vdecomp}), 
the $\mathfrak{sl}(2,\BbbR)$ analysis of 
\cite{bargmann} in each index shows that 
the adjoint relations
\begin{equation}
\bigl({\hat\tau}_0^\eta\bigr)^\dagger =
\bigl({\hat\tau}_0^\eta\bigr)^\dagger
\ \ , \ \ 
\bigl({\hat\tau}_\pm^\eta\bigr)^\dagger =
\bigl({\hat\tau}_\mp^\eta\bigr)^\dagger
\ \ , 
\label{eq:taus-adjoints}
\end{equation}
determine uniquely an inner product in which $\{ \psi_{\mu\nu} \}$ is
an orthonormal set up to an overall scale. This agrees with the inner
product~(\ref{eq:aq-ip}).  The Hilbert spaces
$\mathcal{H}^{\mathrm{e}}$ and
$\mathcal{H}^{\mathrm{o}}_{\epsilon_1\epsilon_2}$ are obtained by Cauchy
completion. 
In the terminology of~\cite{bargmann}, 
the representations of 
$\mathrm{O}_{\mathrm{c}}(2,2) \simeq \bigl( \SLtwor \times \SLtwor
\bigr) /\BbbZ_2$ are respectively 
$C^0_{1/4} \times C^0_{1/4}$
and 
$D^{\epsilon_1}_{1/2} \times D^{\epsilon_2}_{1/2}$. 
The first of these is $T^{1,0}_{2,2}$
and the other four constitute 
$T^{1,1}_{2,2}$. 

On $\mathcal{H}^{\mathrm{e}}$ the spectra of the quantum observables
$\hat{A}_{12} \pm \hat{B}_{12}$ are unbounded both above and
below. This might have been expected on the grounds that the range of
the classical observables ${A}_{12} \pm {B}_{12}$ are unbounded both
above and below. By contrast, on each
$\mathcal{H}^{\mathrm{o}}_{\epsilon_1\epsilon_2}$ the spectra of
$\hat{A}_{12} \pm \hat{B}_{12}$ have a definite sign. That quantum
theories with this property can arise may be related to the failure of
$\Aclass$ to separate the subsets of $\Mreg$ where one of ${A}_{12}
\pm {B}_{12}$ vanishes.

\subsection{$(p,q) = (3,1)$}
\label{subsec:31-low} 

When $(p,q) = (3,1)$, the value of the quadratic Casimir operator
(\ref{eq:pq-quantum-casimir}) in the representations $T^{2,
\epsilon}_{3,1}$ is~$-1$, and a direct computation shows that the Casimir
operator 
$ 
\hat{A}_{12} \hat{C}_{31}
+ 
\hat{A}_{23} \hat{C}_{11} 
+ 
\hat{A}_{31} \hat{C}_{21}
$ has value zero. It follows that the representations $T^{2,
\epsilon}_{3,1}$ are each isomorphic to the principal series unitary
irreducible representation $\mathfrak{S}_{0,0}$ of
$\mathrm{O}_{\mathrm{c}}(3,1)$~\cite{Naimark-book}. Proceeding as in
subsection
\ref{subsec:21-low} yields a decomposition of 
$T^{2,0}_{3,1} \oplus T^{2, 1}_{3,1}$ in which 
each irreducible component is supported on its own branch of the light
cone.

\section{Algebraic quantisation with $\hat{H}_a \Psi=0$}
\label{sec:AQnaive}

In this section we discuss how the algebraic quantisation of section
\ref{sec:AQ} is modified when the constraints
(\ref{eq:amm-Dirac}) are replaced by $\hat{H}_a \Psi=0$. 
We give a complete analysis for
$p+q \le 4$ and partial results for other values of $p$
and~$q$.

\subsection{General $(p,q)$}

The quantum constraints (\ref{eq:qconstraints}) are replaced by 
\begin{subequations}
\label{eq:nqconstraints}
\be
&&\hat{H} \Psi =0
\ \ ,
\label{eq:nqconstraints-H}
\\
&&
\hat{D} \Psi =0
\ \ . 
\label{eq:nqconstraints-D}
\ee
\end{subequations}
Proceeding as in subsection~\ref{subsec:AQ-genpq}, 
the exponent 
of $r$ in (\ref{eq:f-homog}) is replaced by 
$-(p+q-4)/2$, and the value of the quadratic Casimir
(\ref{eq:pq-quantum-casimir}) is $-\frac14 (p+q)(p+q-4)$
\cite{LouMol}. 

The representation of $\mathrm{O}_{\mathrm{c}}(p,q)$ generated by the
quantum observables is now isomorphic to the representation on
homogeneous functions of degree $-(p+q-4)/2$ on the light cone of
$\BbbR^{p,q}$. The outstanding question is whether this representation
or some subrepresentation thereof is unitary in some inner product.

We analyse different ranges of $(p,q)$ in the following 
subsections. By interchange of $p$
and~$q$, it suffices to consider $p\ge q$.

\subsection{$(p,q) = (1,1)$}

When $(p,q) = (1,1)$, the discussion follows 
subsection~\ref{subsec:11-low}. The representation of $\Aphystar$ is
trivial.

\subsection{$(p,q) = (2,1)$}

When $(p,q) = (2,1)$, the factor $u^{-1/2}$ in (\ref{eq:psi-kappa-m})
is replaced by~$u^{1/2}$, and the numerical factor on the right-hand
side of (\ref{eq:21-rep-pm}) is replaced by $\bigl( m \mp \tfrac12
\bigr)$. The representations of $\Aphystar$ on the counterparts of
${\mathcal V}^\kappa$ are irreducible, but there is no inner product
in which $\hat{A}_{12}$, $\hat{C}_{11}$ and $\hat{C}_{21}$ would be
self-adjoint~\cite{bargmann}. No quantum theory is recovered.

\subsection{$(p,q) = (2,2)$}

When $(p,q) = (2,2)$, (\ref{eq:psi22}) is replaced by 
\be
{\tilde\psi}_{\mu \nu}
:= {(-i)}^{\mu + \nu} 
\delta(u^2-w^2)
\, 
e^{i[(\mu+\nu)\alpha +(\mu-\nu)\beta]} 
\ \  ,   
\label{eq:npsi22}
\ee
where $\mu$ and $\nu$ are again either both integers or both
half-integers. We write 
${\tilde{\mathcal{V}}} := 
\mathrm{span} \bigl\{ {\tilde\psi}_{\mu\nu} \bigr\}$. 
A~direct computation gives 
\be
{\hat\tau}_0^+ {\tilde\psi}_{\mu\nu}  
& = & 
\mu {\tilde\psi}_{\mu\nu} 
\ \ , 
\nonumber
\\
{\hat\tau}_\pm^+ {\tilde\psi}_{\mu\nu}  
& = & 
\mu 
{\tilde\psi}_{\mu \pm 1, \nu} 
\ \ , 
\nonumber
\\
{\hat\tau}_0^{-} {\tilde\psi}_{\mu\nu}  
& = & 
\nu {\tilde\psi}_{\mu\nu} 
\ \ , 
\nonumber
\\
{\hat\tau}_\pm^{-} {\tilde\psi}_{\mu\nu}  
& = & 
\nu 
{\tilde\psi}_{\mu, \nu  \pm 1} 
\ \ . 
\label{eq:nrepr22}
\ee
Hence ${\tilde{\mathcal{V}}}$ carries a
representation of~$\Aphystar$. 

We decompose 
${\tilde{\mathcal{V}}}$ 
as ${\tilde{\mathcal{V}}}  = {\tilde{\mathcal{V}}}^{\mathrm{o}}
\oplus 
\bigoplus_{\epsilon_1 \epsilon_2}
{\tilde{\mathcal{V}}}^{\mathrm{e}}_{\epsilon_1 \epsilon_2}$, where 
\be
{\tilde{\mathcal{V}}}^{\mathrm{o}}
&:=&
\mathrm{span}
\bigl\{ {\tilde\psi}_{\mu\nu} 
\mid 
\mu, \nu \in \BbbZ + \tfrac12
\bigr\}
\ \ ,
\nonumber
\\
{\tilde{\mathcal{V}}}^{\mathrm{e}}_{\epsilon_1 \epsilon_2}
&:=&
\mathrm{span}
\bigl\{ {\tilde\psi}_{\mu\nu} 
\mid 
\mu, \nu \in \BbbZ ; \ 
\mathrm{sgn}(\mu) = \epsilon_1 ; \ 
\mathrm{sgn}(\nu) = \epsilon_2
\bigr\}
\ \ ,
\label{eq:n22Vdecomp} 
\ee
and $\epsilon_i \in \{1,0,-1\}$. Equations (\ref{eq:nrepr22}) show
that each of the ten spaces in (\ref{eq:n22Vdecomp}) carries a
representation of~$\Aphystar$, given by (\ref{eq:nrepr22}) except that
whenever a raising (respectively lowering) operator raises (lowers)
the index $-1$ ($+1$) to zero, the vector on the right-hand side is
replaced by the zero vector. It can be verified as in subsection
\ref{subsec:22-low} that these representations are irreducible. The
$\sltwor$ analysis of
\cite{bargmann} in each index then shows that there is no inner product on 
${\tilde{\mathcal{V}}}^{\mathrm{o}}$ compatible with the adjoint
relations~(\ref{eq:taus-adjoints}), whereas on each
${\tilde{\mathcal{V}}}^{\mathrm{e}}_{\epsilon_1 \epsilon_2}$ these
relations determine an inner product that is unique up to an overall
scale.  For $\epsilon_1\ne0\ne\epsilon_2$, this inner product reads
$\bigl( {\tilde\psi}_{\mu'\nu'} , {\tilde\psi}_{\mu\nu} \bigr) = |\mu
\nu| \delta_{\mu \mu'} \delta_{\nu \nu'}$, while the formulas for
$\epsilon_1 \ne 0 = \epsilon_2$ and $\epsilon_1 = 0 \ne \epsilon_2$
are respectively $\bigl( {\tilde\psi}_{\mu' 0} , {\tilde\psi}_{\mu 0}
\bigr) = |\mu| \delta_{\mu \mu'}$ and $\bigl( {\tilde\psi}_{0\nu'} ,
{\tilde\psi}_{0\nu} \bigr) = |\nu| \delta_{\nu \nu'}$. On the
one-dimensional space ${\tilde{\mathcal{V}}}^{\mathrm{e}}_{00}$ the
representation of $\Aphystar$ is trivial, and we have $\bigl(
{\tilde\psi}_{00} , {\tilde\psi}_{00} \bigr) =1$.

The Hilbert spaces 
${\tilde{\mathcal{H}}}^{\mathrm{e}}_{\epsilon_1 \epsilon_2}$ are obtained
by Cauchy completion. In the terminology of~\cite{bargmann}, the
representation of 
$\mathrm{O}_{\mathrm{c}}(2,2) \simeq \bigl( \SLtwor \times \SLtwor
\bigr) /\BbbZ_2$ on 
${\tilde{\mathcal{H}}}^{\mathrm{e}}_{\epsilon_1 \epsilon_2}$
is $D^{\epsilon_1}_1 \times
D^{\epsilon_2}_1$ for $\epsilon_1\ne0\ne\epsilon_2$, and for 
$\epsilon_i=0$, $D^{\epsilon_i}_1$ is
replaced by the trivial representation. 

The quantum theories on
${\tilde{\mathcal{H}}}^{\mathrm{e}}_{\epsilon_1 \epsilon_2}$ with
$\epsilon_1\ne0\ne\epsilon_2$ are qualitatively similar to the
theories on ${\mathcal{H}}^{\mathrm{o}}_{\epsilon_1 \epsilon_2}$
obtained in subsection~\ref{subsec:22-low}, with the roles of integer
and half-integer eigenvalues of ${\hat{\tau}}_0^\eta$
interchanged. The remaining five theories are degenerate in that at
least one ${\hat{\tau}}_0^\eta$ annihilates the whole space. Note that
in contrast to subsection~\ref{subsec:22-low}, we now obtained no
theory in which the representation of $\Aphystar$ would be irreducible
and the spectra of $\hat{A}_{12}
\pm
\hat{B}_{12}$ would be unbounded both above and below.

\subsection{$(p,q)=(3,1)$}
\label{subsec:npq-lorentz} 

When $(p,q)=(3,1)$, the branches $w_1>0$ and $w_1<0$ of the light cone
decouple and give isomorphic quantum theories. For concreteness, we
consider states whose support is on the $w_1>0$ branch. Isomorphic
theories in which the states have support on both branches and have 
respectively even or odd parity
could be constructed as in subsections
\ref{subsec:21-low} 
and~\ref{subsec:31-low}.

We set 
${\tilde{\mathcal V}}
:=
\mathrm{span} \{ {\tilde\psi}_{l m } \}$, 
\label{eq:tildeV31}
where
\be
{\tilde\psi}_{l m }
:= 
\delta(u^2 - w^2) 
\theta(w_1) 
Y_{lm}
\ \ ,
\label{eq:psi-npqlorentz}
\ee
$Y_{lm}$ are the usual spherical harmonics on unit $S^2$ in
$\boldsymbol{u}$ \cite{Bate} and 
$\theta$ is the Heaviside function as in~(\ref{eq:psi-kappa-m}). 
The action of the operators (\ref{eq:q-obser}) on ${\tilde{\mathcal
V}}$ can be computed from standard properties of the spherical
harmonics 
\cite{Bate,arfken} 
and is displayed in Table~\ref{table:observables}. This shows that 
${\tilde{\mathcal
V}}$ carries a representation of~$\Aphystar$. 

\begin{table}[t]
\be
\hat{A}_{12}
{\tilde\psi}_{lm} 
& = & 
m 
{\tilde\psi}_{lm} 
\nn
\\
\bigl( \hat{A}_{23} \pm i \hat{A}_{31} \bigr) 
{\tilde\psi}_{lm} 
& = & \sqrt{(l\pm m+1)(l\mp m)}
\, 
{\tilde\psi}_{l,m\pm1} 
\nn 
\\
\hat{C}_{31}
{\tilde\psi}_{lm} 
& = &
-il \, 
\sqrt{\frac{(l+m+1)(l-m+1)}{(2l+1)(2l+3)}}
\, 
{\tilde\psi}_{l+1, m} 
\nn
\\
&  & 
+ i (l+1) \, 
\sqrt{\frac{(l+m)(l-m)}{(2l+1)(2l-1)}}
\, 
{\tilde\psi}_{l-1, m} 
\nn
\\
\bigl( \hat{C}_{11} \pm i \hat{C}_{21} \bigr) 
{\tilde\psi}_{lm} 
& = &
\pm i l \, 
\sqrt{\frac{(l \pm m +1) (l \pm m +2)}{(2l+1)(2l+3)}}
\, 
{\tilde\psi}_{l+1,m\pm1} 
\nn
\\
&  &
\pm i (l+1) \, 
\sqrt{\frac{(l \mp m) (l \mp m -1)}{(2l+1)(2l-1)}}
\, 
{\tilde\psi}_{l-1,m\pm1} 
\nn
\ee
\caption{The action of $\Aphystar$ on ${\tilde{\mathcal V}}$
for $(p,q)=(3,1)$.}
\label{table:observables}
\end{table}

We decompose 
${\tilde{\mathcal V}}$ as 
${\tilde{\mathcal V}} = {\tilde{\mathcal V}}_0 \oplus 
{\tilde{\mathcal V}}_+$, where 
\be
{\tilde{\mathcal V}}_0
&:=&
\mathrm{span}
\bigl\{ 
{\tilde\psi}_{00 }
\bigr\}
\ \ ,
\nonumber
\\
{\tilde{\mathcal V}}_+
&:=&
\mathrm{span}
\bigl\{ 
{\tilde{\psi}}_{lm }
\mid 
l >0 
\bigr\}
\ \ . 
\ee
Table \ref{table:observables} shows that $\Aphystar$ is represented
trivially on ${\tilde{\mathcal V}}_0$, while ${\tilde{\mathcal V}}_+$
carries a representation that is as in the Table except that any term
on the right-hand side with the first index taking the value zero is
replaced by the zero vector. Comparison with the infinitesimal
representations of $\mathrm{O}_{\mathrm{c}}(3,1)$
(\cite{Naimark-book}, section 8.3) shows that the representation on
${\tilde{\mathcal V}}_+$ is isomorphic to the principal series
irreducible representation $\mathfrak{S}_{2,0}$, which is unitary
precisely when the inner product is
\begin{equation}
\bigl( {\tilde\psi}_{l' m' } , {\tilde\psi}_{l m } \bigr) 
= 
l(l+1) 
\, 
\delta_{l l'} 
\delta_{ m m'} 
\ \ , 
\end{equation}
up to an overall multiple. The Casimir
operator 
$ 
\hat{A}_{12} \hat{C}_{31}
+ 
\hat{A}_{23} \hat{C}_{11} 
+ 
\hat{A}_{31} \hat{C}_{21}
$
takes value zero.

\subsection{$2\le q \le p$, $(p,q) \ne (2,2)$}
\label{subsec:npq-large} 

When $2\le q \le p$ and  $(p,q) \ne (2,2)$ we set 
${\tilde{\mathcal V}}
:=
\mathrm{span} \{ {\tilde\psi}_{lj k_u k_w} \}$, 
where 
\begin{equation}
{\tilde\psi}_{lj k_u k_w} := 
\delta(u^2 - w^2) 
u^{-(p+q-4)/2} 
\, 
Y_{lk_u}
\bigl({\theta}^{(u)}\bigr)
Y_{jk_w}
\bigl({\theta}^{(w)}\bigr)
\ \ ,
\end{equation}
$l$ and $j$ are non-negative integers 
and $Y_{lk_u}$
(respectively $Y_{jk_w}$)
are the
spherical harmonics on unit $S^{p-1}$ in $\boldsymbol{u}$
($S^{q-1}$ in $\boldsymbol{w}$)~\cite{Vil,Bate}.
The notation for the 
angular coordinates ${\theta}^{(u)}$ and ${\theta}^{(w)}$ and the
spherical harmonics follows~\cite{LouMol}. 
The construction of the spherical harmonics implies that
${\tilde{\mathcal V}}$ carries a representation of~$\Aphystar$. 

We seek a linear subspace 
${\tilde{\mathcal V}}_0 \subset {\tilde{\mathcal
V}}$ with the following properties: 
\begin{enumerate} 
\item 
${\tilde{\mathcal V}}_0 := 
\mathrm{span} \bigl\{ {\tilde\psi}_{lj k_u k_w}
\mid (l,j) \in I
\bigr\}$, where $I$ is some nonempty index set. 
\item 
${\tilde{\mathcal V}}_0$ carries a representation of~$\Aphystar$. 
\item
The generators (\ref{eq:q-obser}) of $\Aphystar$ are self-adjoint in
an inner product of the form
\begin{equation}
\bigl( {\tilde\psi}_{l' j' k_u' k_w'} , 
{\tilde\psi}_{l j k_u k_w} \bigr) 
= K_{lj} 
\, 
\delta_{l l'} \delta_{j j'} 
\delta_{ k_u  k_u'} \delta_{ k_w  k_w'} 
\ \ , 
\label{eq:pqlarge-try-ip}
\end{equation}
where the positive numbers
$K_{lj}$ depend only on $l$ and~$j$. 
\end{enumerate}

By the properties of the spherical
harmonics, the rotation generators $\hat{A}_{ij}$ and $\hat{B}_{ij}$ 
in (\ref{eq:q-obser}) leave ${\tilde{\mathcal V}}_0$ invariant and are
self-adjoint in the inner
product~(\ref{eq:pqlarge-try-ip}). What remains is to examine the boost 
generators~$\hat{C}_{ij}$. 

Let $Y_{l0}$ (respectively $Y_{j0}$) denote the zonal spherical
harmonics, which can be expressed in terms of Gegenbauer polynomials of
argument
$u_p/u$ ($w_q/w$)
\cite{Bate}. 
The recursion relations of the Gegenbauer
polynomials~\cite{magnusetal} allow an explicit computation of the
action of $\hat{C}_{pq}$ on ${\tilde\psi}_{l j 0 0}$. Suppressing the
indices $k_u = k_w =0$, we find
\be
4 i \hat{C}_{pq} {\tilde\psi}_{l j}
&=& 
\bigl[ 
l+j + \tfrac12 (p+q-4)
\bigr] 
\left( 
W_{p,l+1} W_{q,j+1}
\, 
{\tilde\psi}_{l+1, j+1}
- 
W_{p l} W_{q j}
\, 
{\tilde\psi}_{l-1, j-1}
\right) 
\nn
\\
&&
+ 
\bigl[ 
l-j + \tfrac12 (p-q)
\bigr] 
\left( 
W_{p,l+1} W_{qj}
\, 
{\tilde\psi}_{l+1, j-1}
- 
W_{p l} W_{q,j+1}
\, 
{\tilde\psi}_{l-1, j+1}
\right) 
\ \ , 
\nn
\\
\label{eq:largepq-Cpq}
\ee
where 
\be
W_{qj}
&:=& 
2\left[\frac{j(j+q-3)}{(2j + q -2)(2j + q - 4)}\right]^{1/2} 
\ \ \ \text{for  $j>0$, $(q,j) \ne (2,1)$}
\ \ , 
\nn
\\
W_{21}
&:=& 
\sqrt{2}
\ \ , 
\nn
\\
W_{q0}
&:=& 
0 
\ \ . 
\label{eq:qW}
\ee
By (\ref{eq:pqlarge-try-ip}) and~(\ref{eq:largepq-Cpq}), 
self-adjointness of 
$\hat{C}_{pq}$ implies the recursion relations 
\begin{subequations}
\label{eq:K-recursion}
\be
\bigl[ 
l+j + \tfrac12 (p+q-4)
\bigr] 
K_{l+1,j+1} 
&=&
\bigl[ 
l+j + \tfrac12 (p+q)
\bigr] 
K_{lj} 
\ \ , 
\label{eq:K-recursion:raise-raise}
\\
\bigl[ 
l-j-1 + \tfrac12 (p-q) 
\bigr] 
K_{l+1,j} 
&=&
\bigl[ 
l-j+1 + \tfrac12 (p-q) 
\bigr] 
K_{l,j+1} 
\ \ .
\label{eq:K-recursion:raise-lower}
\ee
\end{subequations}
Note that the coefficients 
in (\ref{eq:K-recursion:raise-raise}) are
always positive.

Suppose first that $p+q$ is odd. From (\ref{eq:largepq-Cpq}) it follows
that the index set $I$ must contain all pairs $(l,j)$ where $l+j$
is odd or all pairs where $l+j$ is even. The coefficients in
(\ref{eq:K-recursion:raise-lower}) are always nonzero, but the coefficients
on the two sides have opposite sign for $j-l = \tfrac12(p-q \pm1)$. 
Hence there are both positive and negative~$K_{lj}$, and the inner
product does not exist. We have proved: 
\begin{theorem}
Let $p\ge2$, $q\ge2$ and $p+q \equiv 1 \pmod2$. Then there is no
${\tilde{\mathcal V}}_0$ satisfying 1--3. 
\end{theorem}

Suppose then that $p+q$ is even. If ${\tilde{\mathcal V}}_0$ contains
a vector for which $l-j + \tfrac12(p-q)$ is odd,
(\ref{eq:largepq-Cpq}) shows that it must contain all such vectors,
and examination of the signs in (\ref{eq:K-recursion:raise-lower})
shows that the inner product does not exist. Hence ${\tilde{\mathcal
V}}_0$ can contain only vectors for which $l-j + \tfrac12(p-q)$ is
even. From (\ref{eq:largepq-Cpq}) and (\ref{eq:K-recursion}) we
obtain:
\begin{theorem}
\label{theorem:pex}
Let $p\ge2$, $q\ge2$, $(p,q)\ne (2,2)$ and $p+q \equiv 0 \pmod2$. 
For a ${\tilde{\mathcal V}}_0$ satisfying 1--3, $I$ is 
either 
$I_0 := 
\bigl\{
(l,j)
\mid 
l-j + \tfrac12(p-q) = 0
\bigr\}$ 
or one of 
$I_{\pm}
:= 
\bigl\{
(l,j)
\mid 
l-j + \tfrac12(p-q) = 2k, \ k \in \BbbZ_\pm
\bigr\}$ 
or a union of two or all of these. For indices within each of 
$I_0$, $I_+$ and~$I_-$, 
(\ref{eq:K-recursion})
determines the positive numbers $K_{lj}$
uniquely up to an overall multiple. 
\end{theorem}

For $p+q$ even, Theorem \ref{theorem:pex} severely restricts the
possible candidates for~${\tilde{\mathcal V}}_0$. A~more complete
understanding of the $\hat{C}_{ij}$ action would be required to
determine whether the candidate spaces of Theorem
\ref{theorem:pex} indeed have 
properties 2 and~3. It is known (\cite{Vil}, Section 9.2.10) that
there exist subspaces of $\tilde{\mathcal{V}}$ carrying an irreducible
representation of $\Aphystar$ such that the generators
(\ref{eq:q-obser}) are self-adjoint in an appropriate inner product,
and these representations are equivalent to certain discrete series
representations of $\mathrm{O}_{\mathrm{c}}(p,q)$. We shall not
examine correspondences between these subspaces and the spaces
${\tilde{\mathcal V}}_0$ of Theorem~\ref{theorem:pex} here.

\subsection{$p\ge4$, $q=1$}
\label{subsec:npq-glorentz}

When $p\ge4$ and $q=1$, the branches $w_1>0$ and $w_1<0$ of the light
cone decouple and lead to isomorphic situations. We consider the
branch~$w_1>0$. We set ${\tilde{\mathcal V}} :=
\mathrm{span} \{ {\tilde\psi}_{l k_u } \}$, 
where
\be
{\tilde\psi}_{l k_u }
:= 
\delta(u^2 - w^2) 
\theta(w_1) 
u^{-(p-3)/2} 
\,
Y_{l k_u}
\ \ , 
\ee
$\theta$ is the Heaviside function as in (\ref{eq:psi-kappa-m}) and 
(\ref{eq:psi-npqlorentz}) and 
$Y_{l k_u}$ are the spherical harmonics on unit $S^{p-1}$ in
$\boldsymbol{u}$ as in subsection~\ref{subsec:npq-large}. 
We seek an inner product of the form 
$\bigl( {\tilde\psi}_{l' k_u'} , 
{\tilde\psi}_{l k_u} \bigr) 
= K_{l} 
\, 
\delta_{l l'} 
\delta_{ k_u  k_u'} 
$, where the positive numbers $K_{l}$ depend only on~$l$. 
From the action of $\hat{C}_{p1}$ on the $k_u=0$ states 
it then follows as in subsection 
\ref{subsec:npq-large} that a necessary condition for the operators 
(\ref{eq:q-obser}) to be self-adjoint is $K_l = r 
\bigl[ l + \tfrac12 (p-1) \bigr] 
\bigl[ l + \tfrac12 (p-3) \bigr]$, where $r$ is a positive constant. 
This fixes the candidate inner product uniquely up to an overall
multiple. The rotation generators $\hat{A}_{ij}$ are clearly
self-adjoint in this inner product, but further analysis would be
required to see whether the same holds for all the boost
generators~$\hat{C}_{i1}$.

\section{Refined algebraic quantisation with 
a nonunimodular gauge group}
\label{sec:RAQoutline}

In this section we give a brief outline of refined algebraic
quantisation (RAQ) with group averaging when the gauge group is a
connected but not necessarily unimodular Lie group. We follow
\cite{GM2} but include the possibility of non-unimodularity at the
outset and 
take the opportunity to clarify in subsection
\ref{subsec:RAQaveraging} certain technical issues 
that arise from the antilinearity of the RAQ observable action on the 
RAQ physical Hilbert space, on comparison with quantisations in which
the action of the observables on the physical states
is linear. 
A~recent status report of refined algebraic
quantisation can be found in~\cite{Marolf-MG}.

\subsection{Refined algebraic quantisation}
\label{subsec:RAQgeneral}

RAQ begins by choosing an auxiliary Hilbert space
$\Haux$ and implementing the quantum constraints as 
self-adjoint operators on it, such that the 
commutators of the constraints close as a 
Lie algebra and the constraints exponentiate into a unitary 
representation $U$ of a corresponding
connected Lie group~$G$. 
We refer to $G$ as the gauge group of the
quantum theory. We denote the left and right Haar measures on $G$ by
respectively $d_L g$ and~$d_R g$, and we denote their geometric
average by~$d_0 g$. These measures are related by 
$d_0g
= {[\Delta(g)]}^{1/2}  d_L g = {[\Delta(g)]}^{-1/2}  d_R g$, where 
$\Delta(g) := \det ( \mathrm{Ad}_g )$ is the modular function.

We wish to solve the constraints in an enlargement of~$\Haux$. We
introduce a space of test states, a dense linear subspace
$\Phi\subset\Haux$ such that the operators $U(g)$ map $\Phi$ to
itself. The desired enlargement is the algebraic dual of~$\Phi$,
denoted by $\Phi^*$ and topologised by the topology of pointwise
convergence. For $f\in\Phi^*$ and $\phi\in\Phi$, we denote the dual
action of $f$ on $\phi$ by~$f[\phi]$. $\Phi^*$~carries a
representation $U^*$ of $G$ defined by the dual action: For
$f\in\Phi^*$, $\bigl(U^*(g)f\bigr)[\phi] = f[U(g^{-1})\phi]$ for all
$\phi \in
\Phi$. 
Solutions to the quantum constraints are then defined to be the
elements $f\in \Phi^*$ for which $U^*(g)f = {[\Delta(g)]}^{1/2}f$ for
all $g\in G$. We return to the reason for including the factor
${[\Delta(g)]}$ in subsection~\ref{subsec:RAQaveraging}.

The RAQ observable algebra is determined by the above structure. An
operator ${\cal O}$ on $\Haux$ is called gauge invariant if the
domains of ${\cal O}$ and ${\cal O}^\dag$ include~$\Phi$, ${\cal O}$
and ${\cal O}^\dag$ map $\Phi$ to itself, and ${\cal O}$ commutes with
the $G$-action on~$\Phi$. Note that if ${\cal O}$ is gauge invariant,
so is~${\cal O}^\dag$. The algebra of gauge invariant operators is
called the observable algebra and denoted by~$\Aobs$. $\Aobs$~has on
$\Phi^*$ an antilinear representation defined by the dual
action~\cite{GM1}: For $f\in\Phi^*$, $({\cal O}f)[\phi] := f [{\cal
O}^\dag
\phi]$ for all $\phi \in \Phi$. 


The last ingredient in RAQ is a rigging map, which is an antilinear
map $\eta: \Phi \to \Phi^*$ satisfying four postulates:

(i)~The image of $\eta$ solves the constraints. 

(ii)~$\eta$ is real: 
$\eta(\phi_1)[\phi_2] = \overline{\eta(\phi_2)[\phi_1]}$ 
for all $\phi_1, \phi_2 \in \Phi$. 

(iii)~$\eta$ is positive: 
$\eta(\phi)[\phi] \ge 0$
for all $\phi\in \Phi$. 

(iv)~$\eta$ intertwines with the 
representations of the observable algebra on $\Phi$ and~$\Phi^*$: 
${\cal O} ( \eta \phi ) = \eta ({\cal O} \phi)$ for all 
${\cal O}\in\Aobs$ and all $\phi\in\Phi$. 

Now, the RAQ physical Hilbert space $\Hraq$ is defined to be the
completion of the image of $\eta$ in the Hermitian inner product
\begin{equation}
\bigl(\eta(\phi_1), \eta(\phi_2) \bigr)_{\mathrm{RAQ}} 
:= \eta(\phi_2)[\phi_1]
\ \ . 
\label{phys-ip}
\end{equation}
$\Hraq$~carries an antilinear representation of~$\Aobs$, and the
adjoint map in this representation is that induced from the adjoint
map on~$\Haux$.

\subsection{Group averaging}
\label{subsec:RAQaveraging}

The aim of group averaging is to provide a rigging
map. 

We define on $\Phi$ the group averaging sesquilinear form
\begin{equation}
(\phi_2,\phi_1)_\mathrm{ga}
:=
\int_G
d_0 g \,
(\phi_2,U(g)\phi_1)_\mathrm{aux}
\ \ , 
\label{eq:GAM}
\end{equation}
assuming that the integral on the right-hand side converges in
absolute value for all $\phi_1$ and $\phi_2$ in~$\Phi$. The group
averaging rigging map candidate is
\begin{equation}
\eta(\phi_1)[\phi_2] 
:= 
{(\phi_1 , \phi_2 )}_{\rm ga}
\ \ . 
\label{GAM-eta}
\end{equation}
This map satisfies (i) because $d_0 (gh) = {[\Delta(h)]}^{-1/2} d_0
(g)$ for constant $h\in G$ and (ii) because $d_0g =
d_0(g^{-1})$. Property (iv) is clear. If $\eta$ further
satisfies~(iii), and if $\eta$ is not identically zero, the group
averaging rigging map candidate is a rigging map. If the
convergence of the averaging is sufficiently strong, 
in the sense explained in~\cite{GM2}, the 
group
averaging rigging map is the unique map
satisfying the axioms (i)--(iv).

The antilinearity of the rigging map is a technical advantage in
practical computations on the test space, 
but a complication when
one wishes to compare the results to quantisations in which the
constraints and the observables act on physical states linearly as
on~$\Haux$. With the group averaging rigging map, the
antilinear isomorphism required for the comparison is provided by
complex conjugation of the group averaging sesquilinear form. We
denote this antilinear map by $J: \Hraq \to J\bigl(\Hraq\bigr)$. 
Adopting Dirac's bra-ket notation, if $|\phi\rangle$ is in
$\Phi$ and $\langle
\phi|$ is its Hilbert dual state, we have $\langle\!\langle \psi
| :=
\eta(|\phi\rangle) =
\int_G d_0g \, 
\langle \phi| 
U(g)$, where the rightmost expression is understood in the sense of
taking matrix elements from the right with states in~$\Phi$. The gauge
invariance of $\langle\!\langle
\psi |$ reads in this language $\langle\!\langle\psi| U(g) =
\langle\!\langle \psi| {[\Delta(g)]}^{1/2}$. Now, by unitarity of $U$
and the invariance of $d_0g$ under the group inverse, we have 
$| \psi \rangle\!\rangle := J( \langle\!\langle \psi | ) = 
\int_G d_0g \, 
U(g) | \phi \rangle$, where the 
rightmost expression is understood in the sense of
taking matrix elements from the left with states in the Hilbert dual image
of~$\Phi$. The gauge invariance of $| \psi \rangle\!\rangle$ reads
\begin{equation}
U(g) |\psi\rangle\!\rangle =
{[\Delta(g)]}^{-1/2} |\psi\rangle\!\rangle
\ \ , 
\label{eq:U-on-psi}
\end{equation}
and the linear representation of $\Aobs$ on $J\bigl(\eta(\Phi)\bigr)$
is induced from the representation of $\Aobs$ on~$\Phi$. Although
$|\psi\rangle\!\rangle$ may not be in~$\Haux$, in situations where the
generators of $U$ are interpretable as (say) differential operators on
smooth functions on a manifold, the infinitesimal version of
(\ref{eq:U-on-psi}) is well-defined as a differential equation and
given by (\ref{eq:amm-Dirac})\footnote{In~\cite{GM2}, equation (4.2)
should be understood as our~(\ref{eq:U-on-psi}). Consequently,
equation (B6) in
\cite{GM2} should be understood as
$({\mathcal{J}_a}) |\phi\rangle = - (i/2) \mathrm{tr}(\mathrm{ad}_a)
|\phi\rangle$, which agrees with equation (6.4) in \cite{DEGST} and
with our (\ref{eq:amm-Dirac}). We thank Nico Giulini and Don Marolf
for discussions and correspondence on this point.}.

\section{Refined algebraic quantisation for $(p,q) \ne (1,1)$}
\label{sec:RAQus}

In this section we carry out refined algebraic quantisation of our
system for $(p,q) \ne (1,1)$. 
The special case $p=q=1$ will be treated 
in section~\ref{sec:RAQ-low}.

\subsection{Auxiliary Hilbert space and 
representation of the gauge group}
\label{subsec:auxhil}

We use the auxiliary Hilbert space $\Haux \simeq
L^2(\BbbR^{p+q})$ of square integrable functions $\Psi
(\boldsymbol{u},\boldsymbol{v})$ in the inner product
\begin{equation}
(\Psi_1, \Psi_2)_\mathrm{aux}
:=
\int d^p\boldsymbol{u} \, d^q\boldsymbol{v}
\,
\overline{\Psi_1} \Psi_2
\ \ . 
\end{equation}
The algebra of the quantum constraints (\ref{eq:con-op}) exponentiates
to a unitary representation $U$ of $G$ on~$\Haux$. In the
decomposition (\ref{eq:iwasawa-decomp}) of appendix~\ref{app:affine},
the group elements are represented by
\begin{subequations}
\label{eq:Iwasawa}
\be
U\bigl(\exp(\mu e^-)\bigr)
& = &
\exp(-i\mu\hat{H})
\ \ ,
\label{eq:Iwasawa1}
\\
U\bigl(\exp(\lambda h)\bigr)
& = &
\exp(-i\lambda \hat{D})
\ \ , 
\label{eq:Iwasawa2}
\ee
\end{subequations}
where 
\begin{subequations}
\label{eq:iwaop-two-action}
\be
[ \exp (-i\mu\hat{H})  \Psi ] (\boldsymbol{u},\boldsymbol{w})
& = &
\exp \! \left[\frac{i\mu
\left(\boldsymbol{u}^2 - \boldsymbol{w}^2
\right) }{2} 
\right] 
\Psi(\boldsymbol{u},\boldsymbol{w})
\ \ ,
\phantom{aaaa}
\label{eq:iwaop-two-action1}
\\
{[\exp (-i\lambda\hat{D}) \Psi]} (\boldsymbol{u},\boldsymbol{w})
& = &
\exp \! \left[-\frac{\lambda(p+q)}{2} \right]
\Psi(e^{-\lambda}\boldsymbol{u},e^{-\lambda} \boldsymbol{w})
\ \ .
\label{eq:iwaop-two-action2}
\ee
\end{subequations}

\subsection{Test space}
\label{subsec:test-space}

Let 
\begin{equation}
\Psi_{ljmnk_uk_w} (\boldsymbol{u},\boldsymbol{w})
:=
u^{l+2m} w^{j+2n}
e^{-\frac{1}{2}(u^2+w^2)} Y_{lk_u}
\bigl({\theta}^{(u)}\bigr)
Y_{jk_w}
\bigl({\theta}^{(w)}\bigr)
\ \ ,
\label{eq:testfun}
\end{equation}
where $l$, $j$, $m$ and $n$ are non-negative integers and the
spherical harmonics are as in subsection~\ref{subsec:npq-large}.
$p=1$ is covered as the special case in which ${\theta}^{(u)} := u_1 /
u \in
\{1,-1\}$,
$l \in \{0,1\}$,
the index $k_u$ takes only a single value and can be dropped 
and $Y_l \bigl({\theta}^{(u)}\bigr) :=
\bigl({\theta}^{(u)}\bigr)^l / \sqrt{2}$, and similarly for
$q=1$. 

We set $\Phi_0 := \mathrm{span} \{ \Psi_{ljmnk_uk_w} \} =
\left\{ P(\boldsymbol u, \boldsymbol w)
\exp\left[-\tfrac12 (u^2 + w^2) \right]
\mid P(\boldsymbol u, \boldsymbol w) \right.$ 
is a polynomial in $\{ u_i \}$ and 
$\{ w_i \}${}$\left. 
\vphantom{\left[-\tfrac12 (u^2 + w^2) \right]} \right\}$. 
$\Phi_0$ is dense in $\Haux$ and
mapped to itself 
by~$\Aphystar$. We adopt as our test space $\Phi$ the closure of
$\Phi_0$ under the algebra generated by $\{ U(g)
\mid g\in G \}$. 


\subsection{$(\cdot\,,\cdot)_\mathrm{ga}$}
\label{subsec:ga-ip}

We parametrise the elements in $G$ as in (\ref{eq:iwasawa-decomp}) and
normalise the symmetric measure so that $d_0 g = e^{\lambda} d\lambda
d\mu$. We shall show that the group averaging sesquilinear form
$(\cdot\,,\cdot)_\mathrm{ga}$ (\ref{eq:GAM}) is well defined and
evaluate it explicitly.

From 
(\ref{eq:iwasawa-decomp}), (\ref{eq:Iwasawa}) and
(\ref{eq:iwaop-two-action}) we find
\be
[U(g)\Psi_{ljmnk_uk_w}] (\boldsymbol{u},\boldsymbol{w})
&=& 
z^{-[\frac14(p+q) + \frac12(l+j)  + m + n]} 
u^{l+2m} w^{j+2n}
Y_{lk_u}
\bigl({\theta}^{(u)}\bigr)
Y_{jk_w}
\bigl({\theta}^{(w)}\bigr)
\nn
\\
&& 
\ 
\times 
\exp\left[
- \frac{1}{2}
\left( \frac{1}{z} - i \mu\right) u^2 
- \frac{1}{2}
\left( \frac{1}{z} + i \mu\right) w^2 
\right]
\ \ , 
\label{eq:uPsi}
\ee
where $z := e^{2\lambda}$. In $(\Psi_{l'j'm'n'k_u'k_w'},
U(g)\Psi_{ljmnk_uk_w})_\mathrm{aux}$, the angular integrals give the
factor $
\delta_{l l' }
\delta_{j j' }
\delta_{k_u k'_u}
\delta_{k_w k'_w}
$ 
and the integrals over $u$ and $w$ give 
Gamma-functions~\cite{GradRhyz}, with the result 
\be
&&
(\Psi_{l'j'm'n'k_u'k_w'}, U(g) \Psi_{ljmnk_uk_w})_\mathrm{aux}
\nn
\\
&& 
\ \ \ 
= \ 
\delta_{l l' }
\delta_{j j' }
\delta_{k_u k'_u}
\delta_{k_w k'_w}
2^{\frac12 (p+q) + l + j + m + m' + n + n' - 2} 
\Gamma \bigl( \tfrac12 p + l + m + m' \bigr) 
\Gamma \bigl( \tfrac12 q + j + n + n' \bigr) 
\nn
\\
\noalign{\smallskip}
&& 
\ \ \ \ \ \ \ \ 
\times 
\frac{z^{\frac14(p+q) + \frac12 (l+j) + m' + n'}}
{(1-i\omega)^{\frac12p + l + m + m'} 
(1+i\omega)^{\frac12q + j + n + n'}
(1+z)^{\frac12 (p+q) + l + j + m + m' + n + n' }}
\ \ , 
\label{eq:PsiUPsi}
\ee
where we have written $\mu =\omega (1+z)/z$. An elementary analysis
shows that necessary and sufficient conditions for (\ref{eq:PsiUPsi})
to be integrable in absolute value in the measure $d_0 g = \frac12
z^{-3/2}(1+z) d\omega dz$ are
\be
&&
\tfrac14 (p+q -2) + \tfrac12(l+j) + m + n >0
\ \ , 
\nn
\\
&&
\tfrac14 (p+q -2) + \tfrac12(l+j) + m' + n' >0
\ \ .
\label{eq:conv-cond}
\ee
As $p+q>2$ by assumption, (\ref{eq:conv-cond}) is satisfied, and the
integral of (\ref{eq:PsiUPsi}) in the measure $d_0 g$ is well
defined. To evaluate this integral, we perform first the $dz$ integral
by 3.194.3 in~\cite{GradRhyz}. If one of the exponents in the
remaining $d\omega$ integral is less than unity, a contour deformation
brings the integral to a form to which 3.194.3 in \cite{GradRhyz}
applies, and an analytic continuation in the exponents shows that the
result is valid also for larger exponents.  Collecting, we find
\be
&&
(\Psi_{l'j'm'n'k_u'k_w'}, \Psi_{ljmnk_uk_w})_\mathrm{ga}
\ = \ 
\frac{\pi}{2}
\delta_{l l' }
\delta_{j j' }
\delta_{k_u k'_u}
\delta_{k_w k'_w}
\nn
\\
&& 
\ \ \ \ \ \ \ \ 
\times 
\Gamma \bigl( \tfrac14 (p+q-2) + \tfrac12(l+j) + m + n \bigr) 
\Gamma \bigl( \tfrac14 (p+q-2) + \tfrac12(l+j) + m' + n' \bigr) 
\ \ . 
\nn
\\
\label{eq:ga-eval-Phinought}
\ee
$(\cdot\,,\cdot )_\mathrm{ga}$ is hence well defined on $\Phi_0$ and
given by~(\ref{eq:ga-eval-Phinought}). From the relations \cite{GM2}
$d_0 (h g) = {[\Delta(h)]}^{1/2} d_0 g$ and $d_0 (g h) =
{[\Delta(h)]}^{-1/2} d_0 g$ it follows that $(\cdot\,,\cdot
)_\mathrm{ga}$ is well defined on all of $\Phi$ and given by
(\ref{eq:ga-eval-Phinought}) and
\begin{equation}
\bigl( \phi_1, U(g) \phi_2 \bigr)_\mathrm{ga}
= 
\bigl( U(g) \phi_1,  \phi_2
\bigr)_\mathrm{ga}
= 
{[\Delta(g)]}^{1/2}
(\phi_1 , \phi_2 )_\mathrm{ga}
\ \ . 
\label{eq:ga-eval-Phi}
\end{equation}

\subsection{Rigging map}
\label{subsec:rigging}

If follows from (\ref{eq:ga-eval-Phinought}) that the map $\eta$
defined by (\ref{GAM-eta}) has nontrivial image. We now give an
explicit characterisation of the image of $\eta$ and evaluate $(\cdot
\, , \cdot)_\mathrm{RAQ}$, showing that this sesquilinear form defines
an inner product. 

By (\ref{eq:ga-eval-Phi})
we have 
$\eta(\phi_1) [ U(g) \phi_2]
=
\eta\bigl(U(g)\phi_1\bigr) [\phi_2]
= {[\Delta(g)}]^{1/2}\eta(\phi_1) [\phi_2]$, 
and hence 
$\eta \bigl(U(g)\phi\bigr) = 
{[\Delta(g)}]^{1/2}\eta(\phi)$. 
It therefore suffices to
evaluate $\eta(\phi_1) [\phi_2]$ for $\phi_1,\phi_2 \in \Phi_0$.

Let 
\be
\chi_{ljk_uk_w} 
(\boldsymbol{u},\boldsymbol{w}) 
:= 
\delta(u^2 - w^2) 
\, 
u^{-(p+q-2)/2} 
Y_{lk_u}
\bigl({\theta}^{(u)}\bigr)
Y_{jk_w}
\bigl({\theta}^{(w)}\bigr)
\ \ . 
\label{eq:chi-def}
\ee
We interpret $\overline{\chi_{ljk_uk_w}}$ as an element of~$\Phi^*$,
acting on states $\phi \in \Phi$ by
\begin{equation}
\overline{\chi_{ljk_uk_w}} [\phi] 
= 
\int 
d^p \boldsymbol{u} 
\, d^q \boldsymbol{w}
\, 
\overline{\chi_{ljk_uk_w}
(\boldsymbol u, \boldsymbol w)} 
\phi(\boldsymbol{u}, \boldsymbol{w})
\ \ . 
\label{eq:chi-dual-action}
\end{equation}
An explicit computation gives 
\begin{equation}
\overline{\chi_{l'j'k_u'k_w'}} [\Psi_{ljmnk_uk_w}] 
= 
\frac14 
\delta_{l l' }
\delta_{j j' }
\delta_{k_u k'_u}
\delta_{k_w k'_w}
\Gamma \bigl( \tfrac14 (p+q-2) + \tfrac12(l+j) + m + n \bigr) 
\ \ . 
\label{eq:chibar-on-Psi}
\end{equation}
From (\ref{GAM-eta}), 
(\ref{eq:ga-eval-Phinought})
and 
(\ref{eq:chibar-on-Psi}) 
we find 
\be
\eta(\Psi_{ljmnk_uk_w}) 
= 
2 \pi 
\Gamma \bigl( \tfrac14 (p+q-2) + \tfrac12 (l+j) +m+n \bigr) 
\, 
\overline{\chi_{ljk_uk_w}}
\ \ . 
\label{eq:eta-of-Psi}
\ee
Hence the image of $\eta$ is spanned by $\{ \overline{\chi_{ljk_uk_w}}
\}$. From (\ref{phys-ip}), 
(\ref{eq:ga-eval-Phinought}), (\ref{eq:chibar-on-Psi}) and
(\ref{eq:eta-of-Psi}) we obtain
\begin{equation}
\bigl(
\,
\overline{\chi_{l'j'k_u'k_w'}} , 
\overline{\chi_{ljk_uk_w}}
\,
\bigr)_{\mathrm{RAQ}} =
{(8\pi)}^{-1}
\delta_{l l' }
\delta_{j j' }
\delta_{k_u k'_u}
\delta_{k_w k'_w}
\ \ .
\label{eq:raq-ip-explicit}
\end{equation}

We see that $(\cdot\,,\cdot )_\mathrm{RAQ}$ is positive definite. It
follows that $\eta$ is a rigging map and the physical Hilbert space
$\Hraq$ is the Cauchy completion of the image of $\eta$ in
$(\cdot\,,\cdot)_\mathrm{RAQ}$.

The representation (\ref{eq:q-obser}) of $\Aphystar$ on $\Haux$ leaves
$\Phi$ invariant and commutes with~$U(g)$, and the star-relation in
this representation coincides with the adjoint map on~$\Haux$. It
follows that $\Aphystar \subset
\Aobs$. Comparison of (\ref{eq:chi-def}) 
and (\ref{eq:chi-dual-action}) to (\ref{eq:Vphys}) shows that the
representation of $\Aphystar$ on $\Hraq$ is antilinearly isomorphic to
the representation of $\Aphystar$ obtained in the algebraic
quantisation in subsection~\ref{subsec:AQ-genpq}.

\section{Refined algebraic quantisation for $p=q=1$}
\label{sec:RAQ-low}

In this section we carry out refined algebraic
quantisation for $p=q=1$. 

When $p=q=1$, the convergence conditions (\ref{eq:conv-cond}) fail to
hold when $l=j=0$ and at least one of the pairs $(m,n)$ and $(m',n')$
equals $(0,0)$. Further, the integral of (\ref{eq:PsiUPsi}) is
unambiguously divergent for $l=l'=j=j'=m=m'=n=n'=0$. We shall remedy
this problem by modifying the test space.

Dropping the 
redundant indices 
$k_u$ and~$k_w$, we write 
\begin{equation}
\phi_{ljmn} (u_1,w_1)
:= 
\Psi_{ljmn00}
= 
u^{l+2m} w^{j+2n}
e^{-\frac{1}{2}(u^2+w^2)} Y_{l}
\bigl({\theta}^{(u)}\bigr)
Y_{j}
\bigl({\theta}^{(w)}\bigr)
\ \ ,
\end{equation}
where $l,j \in \{0,1\}$. We also define 
\be
\psi_{mn}
&:=& 
- 2 \phi_{00,m+1,n+1}
+ (2m+1) \phi_{00,m,n+1}
+ (2n+1) \phi_{00,m+1,n}
\ \ . 
\label{eq:11psi-def}
\\
\Phi_0^\mathrm{mod} 
&:=& 
\mathrm{span}  \bigl(
\{
\phi_{ljmn} \mid  l+j>0 \} \cup \{ \psi_{mn} \} \bigr)
\ \ . 
\label{eq:Phinoughtmod-def}
\ee
An explicit computation shows 
\begin{subequations}
\label{eq:C11-on-11Phi}
\be
\hat{C}_{11} \phi_{00mn} 
&=&
2i 
\bigl(
- \phi_{11mn}
+ m \phi_{11,m-1,n}
+ n \phi_{11,m,n-1}
\bigr)
\ \ , 
\\
\hat{C}_{11} \phi_{01mn} 
&=&
i 
\bigl[
- 2 \phi_{10,m,n+1}
+ 2m \phi_{10,m-1,n+1}
+ (2n+1) \phi_{10mn}
\bigr]
\ \ , 
\\
\hat{C}_{11} \phi_{10mn} 
&=&
i 
\bigl[
- 2 \phi_{01,m+1,n}
+ 2n \phi_{01,m+1,n-1}
+ (2m+1) \phi_{01mn}
\bigr]
\ \ , 
\\
\hat{C}_{11} \phi_{11mn} 
&=&
i 
\psi_{mn}
\ \ , 
\label{eq:C11-lowlow}
\ee
\end{subequations}
where $\hat{C}_{11}$ (\ref{eq:q-obser}) is the single
generator of~$\Aphystar$. Hence $\Phi_0^\mathrm{mod}$
is invariant
under~$\Aphystar$.  
We show in appendix \ref{app:dense} that 
$\Phi_0^\mathrm{mod}$ is dense
in~$\Haux$. 

We denote by $\Phi^\mathrm{mod}$ the closure of $\Phi_0^\mathrm{mod}$
under the algebra generated by $\{ U(g)
\mid g\in G \}$ and adopt $\Phi^\mathrm{mod}$ as the test space. By
construction, $\Phi^\mathrm{mod}$ is invariant under $\Aphystar$. From
(\ref{eq:conv-cond}), (\ref{eq:11psi-def}) and
(\ref{eq:Phinoughtmod-def}) it follows that the integral in
(\ref{eq:GAM}) converges in absolute value, and $(\cdot\,,\cdot
)_\mathrm{ga}$ is hence well defined. The evaluation of
$(\cdot\,,\cdot )_\mathrm{ga}$ for $l+j>0$ proceeds as in
section~\ref{sec:RAQus}, while $(\psi_{m'n'}, \psi_{mn}
)_\mathrm{ga}=0$ by an explicit computation using
(\ref{eq:ga-eval-Phinought}) and~(\ref{eq:11psi-def}), in the index
range where (\ref{eq:ga-eval-Phinought}) is valid. Hence the formulas
of section \ref{sec:RAQus} hold for $l+j>0$, while $\eta$ sends the
whole $l=j=0$ sector of $\Phi^\mathrm{mod}$ to zero.

$\Hraq$ has dimension three. The representation of $\Aphystar$ on 
$\Hraq$ is trivial.

\section{Discussion}
\label{sec:discussion}

In this paper we have discussed refined algebraic quantisation with
group averaging in a constrained Hamiltonian system with unreduced
phase space $\BbbR^{2(p+q)}$, where $p\ge1$ and $q\ge1$, and a
nonunimodular gauge group. The system arose by restricting
the gauge transformations in a system with the unimodular gauge group
$\SLtwor$, previously analysed in~\cite{LouRov,LouMol}, to the 
connected two-dimensional nonunimodular subgroup $G$ that consists of 
lower triangular matrices with positive diagonal elements. 
We obtained a Hilbert
space with a nontrivial representation of the distinguished
$\mathfrak{o}(p,q)$ observable algebra for $(p,q)\ne(1,1)$, which is
precisely the condition for the classical reduced phase space to be a
symplectic manifold up to a singular subset of measure zero. The
representation was found to be the end-point of the maximally
degenerate principal unitary series~\cite{Vil}, with the quadratic
Casimir taking the value $-\frac14 {(p+q-2)}^2$.

By contrast, the reduced phase space of the $\SLtwor$ system is a
symplectic manifold up to a singular subset of measure zero for
$\mathrm{min}(p,q) \ge2$, but group averaging in this system gives a
Hilbert space with a nontrivial representation of the
$\mathfrak{o}(p,q)$ observables only when $\mathrm{min}(p,q)
\ge2$ and $p+q \equiv 0 \pmod2$ \cite{LouRov,LouMol}. The quadratic Casimir
in this representation takes the value $-\frac14 (p+q)(p+q-4)$. The
difference in the Casimirs in the $G$ system and the $\SLtwor$ system
arises from the different senses in which the physical states produced
by group averaging satisfy the constraints for unimodular and
nonunimodular gauge groups~\cite{GM2}. The physical states in the
$\SLtwor$ system are invariant under $\SLtwor$, in a representation
induced from the unitary representation on the auxiliary Hilbert
space, while the physical states in the $G$ system are not invariant
under the corresponding representation of~$G$.

As neither $\SLtwor$ nor $G$ is compact, convergence of the group
averaging was an issue in both systems. We found for each system a
test space on which the convergence is sufficiently strong for the
group averaging formulation of \cite{GM2} to apply, and we could
further choose this space so that the $\mathfrak{o}(p,q)$ observables
are included in the physical observable algebra. In the $\SLtwor$
system the uniqueness theorem of Giulini and Marolf \cite{GM2} implied
that the group averaging rigging map is the only rigging map admitted
by the auxiliary Hilbert space, the representation of the gauge group
and the test space. In the $G$ system the number of rigging maps
admitted by the auxiliary Hilbert space, the representation of the
gauge group and the test space in our $G$ system remains open, even
though the test spaces in the two systems are very similar. The reason
for this difference is that for a nonunimodular gauge group the
uniqueness theorem of Giulini and Marolf assumes absolute convergence
not just in the measure $d_0g$ in which the actual averaging is
performed, but in the whole family of measures $\{ \Delta^{n/2}(g) \,
d_0 g \mid n\in\BbbZ\}$, and for our test space this convergence fails
for large~$|n|$.

We also quantised the nonunimodular gauge system in the algebraic
quantisation scheme of~\cite{Ash1,Ash2}, seeking an inner product in
which the classical $\mathfrak{o}(p,q)$ observables are promoted into
an irreducibly-represented algebra of self-adjoint operators. When we
took the physical states to satisfy the constraints in a sense that
corrresponds to that in group averaging, we found quantum theories
that are isomorphic to the irreducible representations of the
$\mathfrak{o}(p,q)$ algebra in the group averaging quantum theory, and
we displayed these theories explicitly for $p+q \le4$. When we took
the physical states to be strictly annihilated by the generators of
the $G$-action, we found qualitatively similar quantum theories for
$(p,q) = (1,3)$, $(3,1)$ and $(2,2)$, but for $(p,q) = (1,2)$ and
$(2,1)$, and for $p\ge2$, $q\ge2$ and $p+q
\equiv 1
\pmod2$, 
algebraic quantisation produced no quantum theory, under certain
technical assumptions that parallel those of the group averaging
theory. It would be of interest to understand to what extent these
phenomena, in particular our failure to find quantum theories for some
$(p,q)$, result from the technical input we used and to what extent
they arise from deeper reasons in the representation theory of
${\mathrm{O}}(p,q)$. Similar questions arose in algebraic and refined
algebraic quantisation of the $\SLtwor$ system in~\cite{LouMol}.

Our analysis adapts readily to the system in which $G$ is replaced by
the full lower triangular subgroup $\Gext$ of $\SLtwor$. Let
$\mathrm{Id}$ stand for the $2\times2$ identity matrix. $\Gext$
consists of two connected components, the component of identity
is~$G$, $\{\mathrm{Id}, - \mathrm{Id}\}\simeq \BbbZ_2$ is a normal
Abelian subgroup of~$\Gext$, and $\Gext/\{\mathrm{Id}, - \mathrm{Id}\}
\simeq G$. From (\ref{eq:gauge-transf}) it follows that the reduced
phase space of the $\Gext$ system is the quotient of $\Mred$ under the
$\BbbZ_2$ group generated by the involutive map $P: (\boldsymbol{u},
\boldsymbol{w}, \boldsymbol{p}, \boldsymbol{\wmom})
\mapsto 
(-\boldsymbol{u}, -\boldsymbol{w}, -\boldsymbol{p},
-\boldsymbol{\wmom})$. From subsection \ref{subsec:red-phasespace} it
is seen that this quotient acts individually on each of $\Mnought$,
$\Mex$ and~$\Mreg$, the quotient of $\Mreg$ is a manifold with the
induced symplectic structure, and $\Aclass$ separates the quotient of
$\Mreg$ up to a set of measure zero. Two choices to promote $P$ into a
quantum operator are $\hat{P}_\epsilon: \Psi( \boldsymbol{u},
\boldsymbol{w} )
\mapsto 
{(-1)}^\epsilon \Psi( -\boldsymbol{u}, -\boldsymbol{w} )$, 
where the index $\epsilon \in \{0,1\}$ labels the
choice. Given~$\epsilon$, both algebraic and refined algebraic
quantisations are then modified by restriction to parity $\epsilon$
states.

Like previous work \cite{GoMa,LouRov,LouMol}, our work displayed the
dual role of the test space in refined algebraic quantisation. This
space is a fundamental technical ingredient in achieving convergence
of the group averaging with a noncompact gauge group, and it has a
deep physical significance in that it determines the algebra of
physical observables. In specific systems like ours, these issues of
principle become however intertwined with the practical issue of
finding a test space for which the rigging map can be 
analysed and evaluated at a sufficient level of detail. 
General theorems on the existence of test spaces that lead
to physically reasonable quantisations are still very much
lacking. One expects this question to remain significant when
techniques akin to group averaging are extended to systems in which
the constraint algebra closes with nonconstant structure
functions~\cite{Shvedov}.

\section*{Acknowledgements}

We thank Abhay Ashtekar, John Barrett, Nico Giulini, Eli Hawkins, Don
Marolf and Jacek Wi\'sniewski for discussions and correspondence, and
in particular Abhay Ashtekar for raising the question addressed in
section~\ref{sec:AQnaive}. J.L. thanks Nico Giulini for hospitality
funded by DFG project Ro 864/7-2 at the University of
Freiburg. A.M. was supported by a CONACYT (Mexico) Postgraduate
Fellowship and an ORS Award to the University of Nottingham.

\appendix

\section{Appendix: Gauge group}
\label{app:affine}

In this appendix we collect some relevant 
properties of the gauge group~$G$. 
The notation follows the $\SLtwor$ notation of~\cite{Howe}.

$G$~is the lower triangular subgroup of $\SLtwor$ with positive
diagonal elements. $G$~is two-dimensional,
nonabelian and connected, which properties 
characterise $G$ uniquely up to isomorphisms. 

The Lie algebra $\mathfrak{g}$ is spanned by the matrices
\begin{equation}
h :=
\left(
\begin{array}{cc}
1 & 0
\\
0 & -1
\end{array}
\right)
\ \ , \ \
e^- :=
\left(
\begin{array}{cc}
0 & 0
\\
1 & 0
\end{array}
\right)
\ \ ,
\end{equation}
whose commutator is
\begin{equation}
\left[ h \, , \,  e^-  \right]
=
-2e^-
\ \ . 
\label{eq:g-basiscomm}
\end{equation} 
Elements of $G$ can be written uniquely as 
\begin{eqnarray} 
g
&=&
\exp(\mu e^-)
\exp(\lambda h)
\nn
\\
\noalign{\smallskip}
&=&
\left(\begin{array}{cc}
1   & 0 \\
\mu & 1
\end{array}\right)
\left(\begin{array}{cc}
e^\lambda & 0 \\
0         & e^{-\lambda}
\end{array}\right)
\ \ ,
\label{eq:iwasawa-decomp}
\end{eqnarray}
where $\mu\in\BbbR$ and $\lambda\in\BbbR$. 

From (\ref{eq:iwasawa-decomp}) we have $g^{-1} dg = h d\lambda + e^-
e^{2\lambda} d\mu$ and $dg g^{-1} = h d\lambda + e^- (d\mu + 2\mu
d\lambda)$. The left and right invariant Haar measures are thus
respectively $d_L g = e^{2\lambda} d\lambda d\mu$ and $d_R g =
d\lambda d\mu$.

The adjoint action of $G$ on $\mathfrak{g}$ reads $\mathrm{Ad}_g (h) =
g h g^{-1} = h + 2\mu e^-$, $\mathrm{Ad}_g (e^-) = g e^- g^{-1} =
e^{-2\lambda} e^-$. Hence the modular function \cite{GM2} is
$\Delta(g) := \det ( \mathrm{Ad}_g ) = e^{-2\lambda}$.  The symmetric
measure, invariant under $g \mapsto g^{-1}$, is $d_0g =
{[\Delta(g)]}^{1/2} d_L g = {[\Delta(g)]}^{-1/2} d_R g = e^{\lambda}
d\lambda d\mu$.

\section{Appendix: Separation of $\Mreg$ by $\Aclass$}
\label{app:separation}

In this appendix we verify the separation properties of $\Aclass$ on
$\Mreg$ stated in subsection~\ref{subsec:red-phasespace}. We represent
$\Mreg$ as the set $\Nreg$ defined by~(\ref{eq:Mreg-char}).

\begin{theorem}
\label{theorem:Mreg-separation-high}
Let $p>1$ and $q>1$, and let $\Nreg^+$ be the subset of $\Nreg$ where 
$0
\ne
\boldsymbol{p}^2 \ne
\boldsymbol{\wmom}^2
\ne 0$. Then $\Aclass$ separates $\Nreg^+$ up to
the twofold degeneracy~(\ref{eq:separ-degen}). 
\end{theorem}

\myproof
Let 
$a = (\boldsymbol{u},\boldsymbol{p},\boldsymbol{w},\wmomvec) \in
\Nreg^+$ 
and 
$b = (\boldsymbol{u}',\boldsymbol{p}',\boldsymbol{w}',\wmomvec') \in
\Nreg^+$ 
such that $A(a) = A(b)$ for all $A \in \Aclass$. 
We shall show that $a=\pm b$. We use the basis~(\ref{eq:obser}). 

Using (\ref{eq:Mreg-char}), 
the condition $A_{ij}(a) = A_{ij}(b)$ implies 
\begin{subequations}
\label{eq:uprimepprime}
\begin{eqnarray}
\boldsymbol{u}' 
&=& 
\cos(\theta) \boldsymbol{u} + \sin(\theta)
{|\boldsymbol{p}|}^{-1} \boldsymbol{p}
\ \ , 
\\
\boldsymbol{p}' 
&=& 
\cos(\theta) \boldsymbol{p} - \sin(\theta)
|\boldsymbol{p}| \boldsymbol{u} 
\ \ , 
\end{eqnarray}
\end{subequations}
where $0\le \theta < 2\pi$. 
Similarly, the condition 
$B_{ij}(a) = B_{ij}(b)$ implies 
\begin{subequations}
\label{eq:piprimevprime}
\begin{eqnarray}
\boldsymbol{w}' 
&=& 
\cos(\varphi) \boldsymbol{w} + \sin(\varphi)
{|\wmomvec|}^{-1} \wmomvec
\ \ , 
\\
\wmomvec' 
&=& 
\cos(\varphi) \wmomvec - \sin(\varphi)
|\wmomvec| \boldsymbol{w} 
\ \ , 
\label{eq:vprime}
\end{eqnarray}
\end{subequations}
where $0\le \varphi < 2\pi$. With 
(\ref{eq:uprimepprime}) and~(\ref{eq:piprimevprime}), 
the condition $C_{ij}(a) = C_{ij}(b)$ reads
\begin{eqnarray}
0 
&=&
p_i w_j 
\left[
\cos(\theta) \cos(\varphi) 
- 1 
+ \sin(\theta) \sin(\varphi) |\wmomvec| {|\boldsymbol{p}|}^{-1}
\right]
\nn
\\
&&
- 
u_i \wmom_j 
\left[
\cos(\theta) \cos(\varphi)  
- 1
+ \sin(\theta) \sin(\varphi) |\boldsymbol{p}| {|\wmomvec|}^{-1}
\right]
\nn
\\
&&
+
u_i w_j 
\left[
\cos(\theta) \sin(\varphi)  |\wmomvec|
- \sin(\theta) \cos(\varphi)  |\boldsymbol{p}|
\vphantom{{|\boldsymbol{p}|}^{-1}} 
\right]
\nn
\\
&&
+
p_i \wmom_j 
\left[
\cos(\theta) \sin(\varphi)  {|\wmomvec|}^{-1} 
- \sin(\theta) \cos(\varphi)  {|\boldsymbol{p}|}^{-1} 
\right]
\ \ . 
\label{eq:C-separation}
\end{eqnarray}
Contracting (\ref{eq:C-separation}) with $u_i w_j$, $u_i \wmom_j$, 
$p_i w_j$ and $p_i \wmom_j$ shows that each of the four terms must vanish
individually. Using ${|\boldsymbol{p}|} \ne {|\wmomvec|}$, elementary
algebra shows that the only solutions are $\theta=\varphi=0$ and 
$\theta=\varphi=\pi$. 
$\blacksquare$

\myremark
The assumption ${|\boldsymbol{p}|} \ne {|\wmomvec|}$ is necessary, for
otherwise (\ref{eq:C-separation}) is satisfied by $\theta=\varphi$. 

\begin{theorem}
\label{theorem:Mreg-separation-low}
Let $p=1$ and $q>1$ (respectively $p>1$ and $q=1$), 
and let $\Nreg^+$ be the subset of $\Nreg$ where
$\wmomvec^2 \ne 0$ 
($\boldsymbol{p}^2 \ne 0$). 
Then $\Aclass$ separates $\Nreg^+$ up to the
twofold degeneracy~(\ref{eq:separ-degen}). 
\end{theorem}

\myproof
It suffices to consider $p=1$ and $q>1$. 

Proceeding as in the proof of Theorem~\ref{theorem:Mreg-separation-high},
the condition $A_{ij}(a) = A_{ij}(b)$ is an identity, 
the condition 
$B_{ij}(a) = B_{ij}(b)$ leads to~(\ref{eq:piprimevprime}), and the 
condition $C_{ij}(a) = C_{ij}(b)$ reads $u_1 \wmomvec = u_1'
\wmomvec'$. As $(u_1')^2 = (u_1)^2 = 1$, the result follows by
contracting  (\ref{eq:vprime}) with~$\wmomvec$. 
$\blacksquare$

\section{Appendix: $\Phi_0^\mathrm{mod}$ is dense in~$\Haux$} 
\label{app:dense}

\begin{theorem}
For $p=q=1$, the test space
$\Phi_0^\mathrm{mod}$ defined in section \ref{sec:RAQ-low} is dense
in~$\Haux$. 
\end{theorem}


\myproof
Let $\Hauxplusplus \subset \Haux$ be the subspace in which the
functions are even both in $u_1$ and
in~$w_1$. By~(\ref{eq:Phinoughtmod-def}), it suffices to show that
$\Phi_0^{\mathrm{mod}++} := \mathrm{span} \{ \psi_{mn} \}$ is dense
in~$\Hauxplusplus$.

Let 
\be
\tilde{\phi}_{mn} 
:=
2^{-(m+n)/2}  {(\pi m! n!)}^{-1/2}
H_m(u_1)
H_n(w_1) 
\exp \bigl[-\tfrac12 (u_1^2 + w_1^2) \bigr] 
\ \ ,
\ee
where the $H$s are the Hermite polynomials~\cite{magnusetal}. 
$\{ \tilde\phi_{mn} \}$ is an orthonormal basis of~$\Haux$, 
$\{ \tilde\phi_{2m , 2n} \}$ is an orthonormal basis of~$\Hauxplusplus$,
and the recursion relations of the Hermite
polynomials imply 
\be
\hat{C}_{11} 
\tilde{\phi}_{mn} 
=
i 
\left( 
- 
\sqrt{(m+1)(n+1)} 
\, \tilde{\phi}_{m+1,n+1} 
+ 
\sqrt{mn} 
\, \tilde{\phi}_{m-1,n-1} 
\right) 
\ \ . 
\label{eq:C11-tlowlow}
\ee
As 
$\mathrm{span} \{ \phi_{11mn} \} = \mathrm{span} 
\{ \tilde\phi_{2m+1, 2n+1} \}$, 
(\ref{eq:C11-lowlow}) 
shows
that 
$\Phi_0^{\mathrm{mod}++} = \mathrm{span} \{ \hat{C}_{11} \tilde\phi_{2m+1,
2n+1} \}$. 

Suppose now that $\Phi_0^{\mathrm{mod}++}$ is not dense in
$\Hauxplusplus$. Then there exists a nonzero vector $y = \sum_{mn}
a_{mn} \tilde\phi_{2m , 2n} \in \Hauxplusplus$ that is orthogonal to
each $\hat{C}_{11} \tilde\phi_{2m'+1,
2n'+1}$. By~(\ref{eq:C11-tlowlow}), the orthogonality implies
\be
a_{m+1,n+1}
= \sqrt{ \frac{(m+\frac12)(n+\frac12)} {(m+1)(n+1)} } 
\, 
a_{mn}
\ \ ,
\ee
from which it follows 
by elementary analysis that $y$ has finite norm only if it is
the zero vector.  Hence $\Phi_0^{\mathrm{mod}++}$ is dense
in~$\Hauxplusplus$, and $\Phi_0^\mathrm{mod}$ is dense
in~$\Haux$. 
$\blacksquare$


\end{document}